\documentclass[aps,preprint,showkeys,nofootinbib,superscriptaddress]{revtex4-1}

\newtheorem{theorem}{Theorem}[section]

\newtheorem{lemma}[theorem]{Lemma}

\usepackage{hyperref}
\usepackage{color}
\definecolor{darkgreen}{rgb}{0,0.35,0}

\usepackage{amsfonts}
\usepackage{amsmath}
\usepackage{subfig}
\usepackage{mathtools,slashed,tensor}
\usepackage[font=scriptsize,labelfont=bf]{caption}
\usepackage{float}

\newcommand{\ucharles}{Faculty of Mathematics and Physics, Charles University, V Hole\v{s}ovi\v{c}k\'ach 2, 18000 Prague 8, Czech Republic}
\newcommand{\kulak}{KU Leuven Campus Kortrijk - KULAK, Department of Physics, Etienne Sabbelaan 53, 8500 Kortrijk, Belgium}

\begin{document}

\title{(Anti-)de Sitter, Poincar\'{e}, Super symmetries, and \\ the two Dirac points of graphene}

\author{A.~Iorio}
\email{alfredo.iorio@mff.cuni.cz}
\affiliation{\ucharles}
\author{P.~Pais}
\email{pais@ipnp.troja.mff.cuni.cz}
\affiliation{\ucharles}
\affiliation{\kulak}

\begin{abstract}
We propose two different high-energy-theory correspondences with graphene (and related materials) scenarios, associated with grain boundaries, that are topological defects for which both Dirac points are necessary. The first correspondence points to a $(3+1)$-dimensional theory, with nonzero torsion, with spatiotemporal gauge group $SO(3,1)$, locally isomorphic to the Lorentz group in $(3+1)$ dimensions, or to the de Sitter group in $(2+1)$ dimensions. The other correspondence treats the two Dirac fields as an internal symmetry doublet, and it is linked here with unconventional supersymmetry with $SU(2)$ internal symmetry. Our results are suggestive, rather than conclusive, and pave the way to the inclusion of grain boundaries in the emergent field theory picture associated with these materials, whereas disclinations and dislocations have been already well explored.
\end{abstract}

\keywords{Emerging quantum fields and spacetimes; Graphene correspondence; Three-dimensional gravity; Unconventional Supersymmetry}

% \pacs{04.60.-m; 04.70.Dy}

\maketitle

% \tableofcontents

\section{Introduction}
\label{sec:motivationsANDplanofthepaper}

We have learned, first theoretically \cite{semenoff1984,GONZALEZ1993771, GonzalezFullerenePRL} and then experimentally \cite{Novoselov666,PacoReview2009} that graphene, and other materials \cite{Claudia,Jakubsky2016,RevModPhys.90.015001}, realize ``spinors quasi-particles'', i.e., particles whose Dirac or Weyl properties emerge due to the structure of the space (lattice) with which the electrons interact. We have also learned how the emergence of intrinsic as well as extrinsic curvature in graphene can be used to probe the fundamental physics of the quantum Dirac field theory in the presence of a variety of curved, but torsion-free, spacetimes \cite{Iorio:2013ifa, Iorio:2011yz, IorioPais} (see also the review \cite{IorioReview}) and even to probe certain quantum gravity scenarios \cite{Iorio:2017vtw}, (see also \cite{Acquaviva:2017xqi,Acquaviva:2017krr}).

This epistemological approach (see, e.g., \cite{Dardashti2016arXiv160405932D} and \cite{Volovik:2003fe, Barceló2011}) is becoming increasingly popular, for a variety of open theoretical questions on fundamental physics addressed via corresponding systems: from the Hawking phenomenon in Bose-Einstein condensates \cite{Steinhauer:2015saa}, to the interpretation of hadronization in high energy collisions as a Unruh phenomenon \cite{Castorina:2007eb} (see also \cite{Castorina:2008gf} and \cite{Castorina:2014fna}), to the recent work on anomalies of various kinds \cite{Gooth:2017mbd}, or reduced quantum electrodynamics \cite{Dudal:2018mms}.

In this paper we move in this spirit, and probe two scenarios where the peculiar feature of having two Dirac points, hence two copies of the Dirac Hamiltonian, is crucial.

In one approach, that we call the \textit{spatiotemporal}, we explore how certain topological defects, for which it is necessary to have two Dirac points, require an extension of the geometric/relativistic corresponding gauge group from Poincar\'{e}, $ISO(2,1)$, to de Sitter or Anti-de Sitter, $SO(3,1)$ or $SO(2,2)$, respectively. The logic here is that the emergence of a phenomenological length parameter, $R$, related to a certain continuum field description of the topological defect, can be associated to a ``cosmological constant'' $\lambda \sim 1/R^2$, that turns translations into rotations.

In the second approach, we focus on the less unusual scenario that treats the two Dirac points as an internal symmetry doublet, offering here an analysis that links it with the recently proposed unconventional supersymmetry (USUSY) \cite{AVZ}, for the case of an $SU(2)$ internal symmetry \cite{u-susy-su2}. Besides its obvious potential impact on high energy physics, USUSY, that finds its most natural setting in $(2+1)$ dimensions, has the intriguing feature to share many aspects of the physics of the Dirac quasiparticles on curved graphene spacetimes, see, e.g., the recent \cite{Andrianopoli2018}.

\begin{figure}
  \centering
  \includegraphics[width= .7\textwidth]{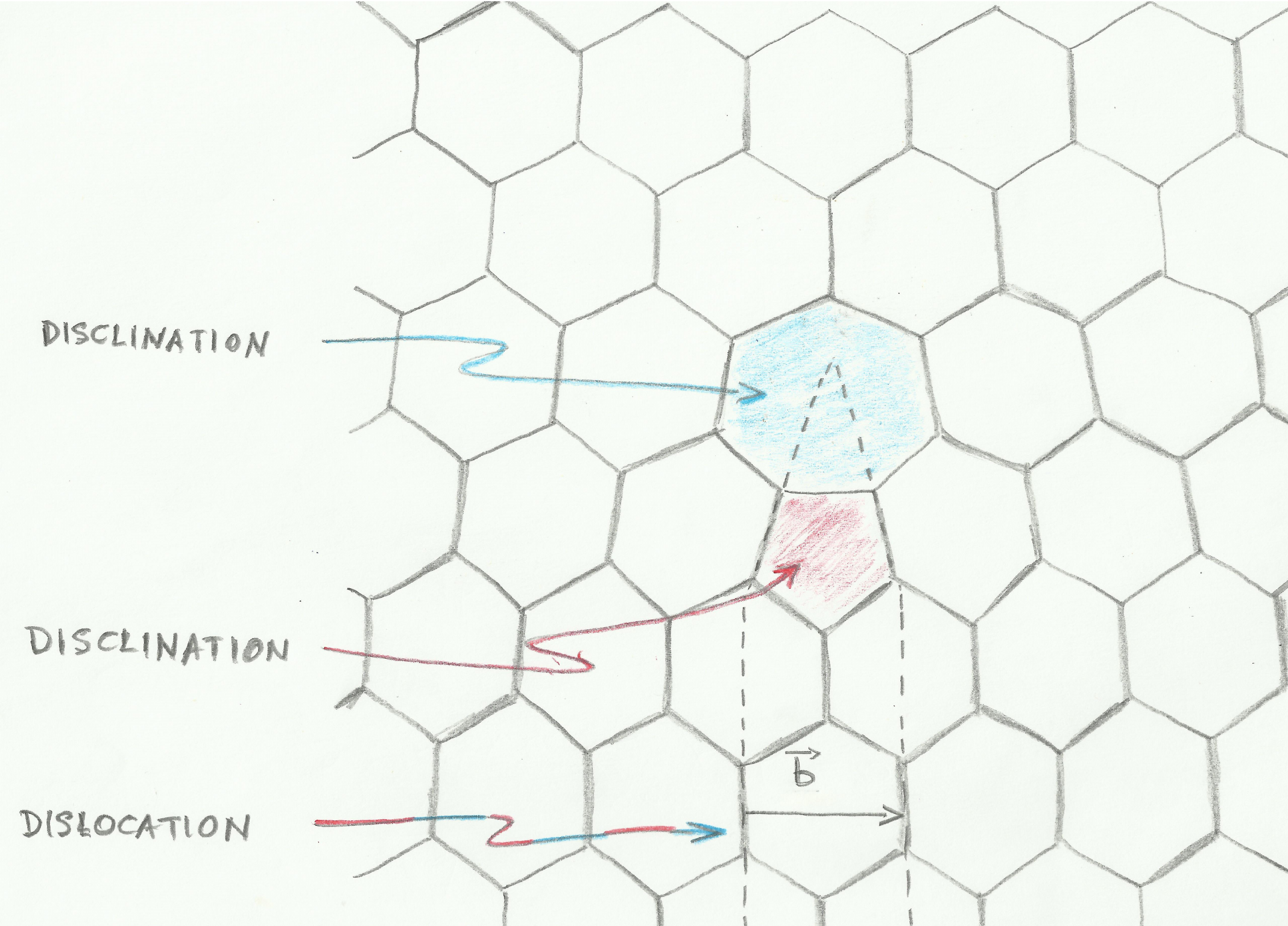}
 \caption{{\bf Edge dislocation from two disclinations}. Two \textit{disclinations}, one carrying one unit of negative intrinsic curvature (the heptagon), one carrying one unit of positive intrinsic curvature (the pentagon), thus adding up to zero total intrinsic curvature but making a \textit{dislocation} with Burger vector $\vec{b}$, carrying torsion.}
  \label{fig:edge-dislocation}
\end{figure}

Before moving to our analysis, let us recall here a very well known story, as this will help us in the following. Notice that, although we have primarily graphene in mind, many of the following considerations apply also to other two-dimensional crystals, with hexagonal symmetry, including, e.g., silicene, germanene, dichalcogenides and the artificial graphene, among others (see \cite{Claudia,Jakubsky2016}).

The two most important topological defects of these materials are \textit{disclinations} and \textit{dislocations}\footnote{Other defects are vacancies, impurities, wrinkles, resonant scatterers, etc.}. Here, and in any other crystal, the two are related to curvature and torsion, respectively, and among themselves, see, e.g., \cite{Kleinert:1989ky}, and later here.
A disclination defect, within a hexagonal lattice, is an $n$-sided polygon with $n = 3, 4, 5$, or $n = 7, 8, 9, ...$. In the first cases, ($n< 6$), the associated conical singularity carries a positive intrinsic curvature, whereas, in the other cases, ($n > 6$), it carries a negative intrinsic curvature. The Frank vector takes that into account, by measuring the deficit ($n < 6$) or excess ($n > 6$) angle, also called disclination angle $s$, and its orientation \cite{Kleinert:1989ky}. In the continuum limit (corresponding, in graphene, to the large wave-length regime of the $\pi$ electrons, when the Dirac description sets in), one can associate to the disclination defect the spin-connection $\omega_\mu^{a b}$ as a gauge field \cite{Kleinert:1989ky,Katanaev:1992kh}. As well known, the corresponding field-strength $F_{\mu \nu} (\omega)$ is the Riemann tensor \cite{Kleinert:1989ky, Katanaev:1992kh}
\begin{equation}\label{Riemann}
    R^{a b}_{\mu \nu} = \partial_{\nu} \; \omega^{a b}_{\mu} + \omega^{a c}_{\nu}\tensor{\omega}{_{c}^{b}_{\mu}} - \nu \leftrightarrow \mu \;,
\end{equation}
with $\tensor{R}{^{\rho}_{\lambda\mu\nu}} = \tensor{E}{^{\rho}_{a}} \tensor{R}{_{\mu\nu}^{a}_{b}} \tensor{e}{^{b}_{\lambda}}$,
$R_{\lambda\nu} = \tensor{E}{^{\mu}_{a}} \tensor{R}{_{\mu \nu}^{a}_{b}} \tensor{e}{^{b}_{\lambda}}$, $R = \tensor{E}{^{\mu}_{a}} \tensor{R}{_{\mu\nu}^{ab}} \tensor{E}{^{\nu}_{b}}$, where $\tensor{e}{^{a}_{\mu}}$ are $\tensor{E}{_{a}^{\mu}}$ the vielbein and its inverse, respectively.

A dislocation defect can be produced by a dipole of disclinations that has zero total curvature (e.g., a $1$-pentagon -- $1$-heptagon pair, a $1$-square -- $1$-octagon pair, a $2$-pentagon -- $1$-octagon pair, etc.) separated by a given distance \cite{Yazyev2010}. The Burger vector $\vec{b}$ takes that into account, by measuring the lack of translation symmetry, as is shown in Fig \ref{fig:edge-dislocation}. In the continuum limit one can associate it to the torsion tensor \cite{Kleinert:1989ky,Katanaev:1992kh}
\begin{equation}\label{Torsion}
\tensor{T}{^{a}_{\mu \nu}} = \partial_{\mu} \; \tensor{e}{^{a}_{\nu}} + \tensor{\omega}{^{a}_{b}_{\mu}} \tensor{e}{^{b}_{\nu}} - \mu \leftrightarrow \nu \;,
\end{equation}
where $\tensor{T}{^{\rho}_{\mu \nu}} = \tensor{E}{_{a}^{\rho}}\tensor{T}{^{a}_{\mu \nu}}$. In other words, as well known, the torsion tensor is the covariant curl of the vielbein.

\begin{figure}
  \centering
  \includegraphics[width= 1\textwidth]{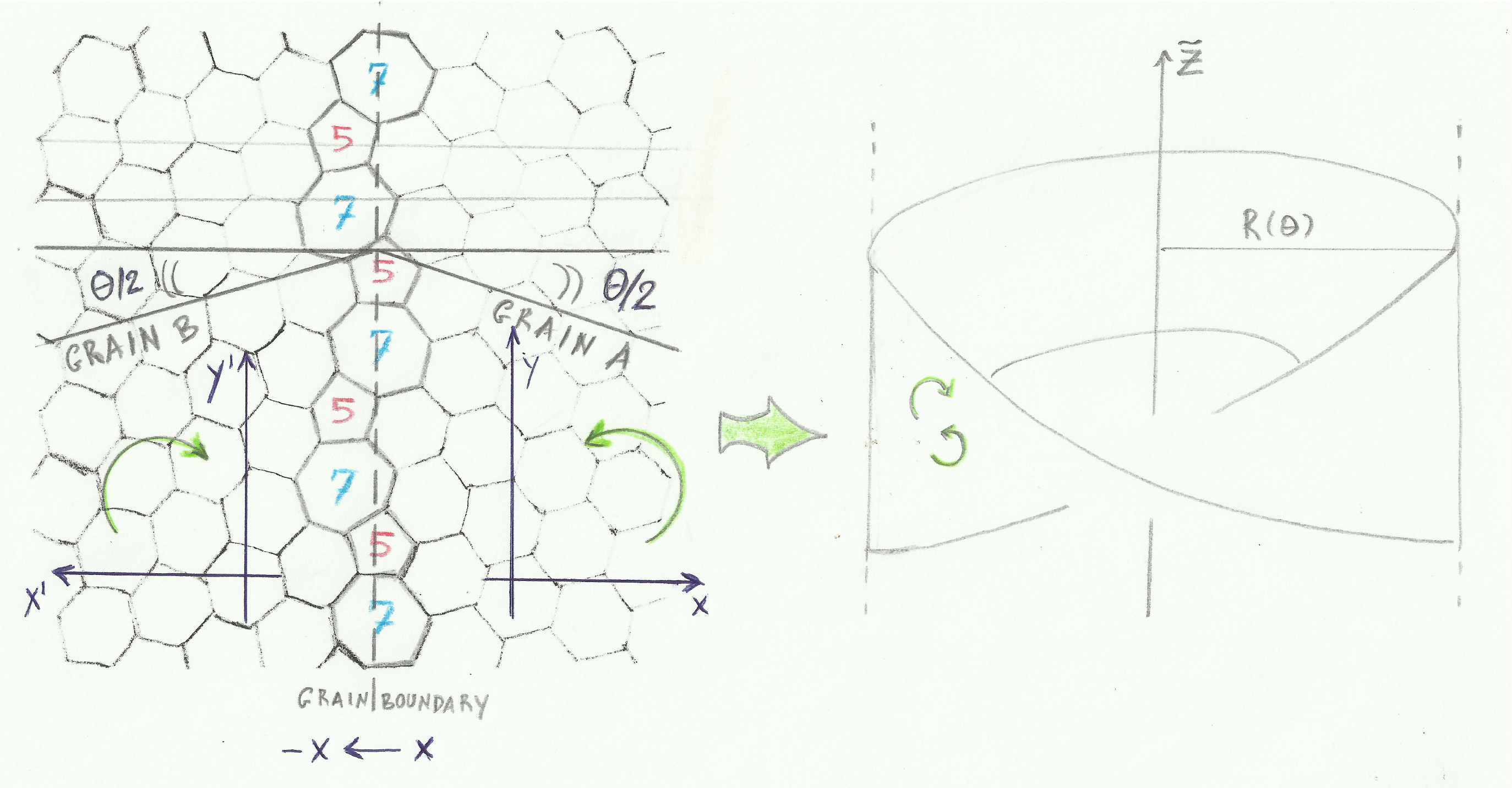}
 \caption{{\bf A grain boundary (left), and a possible modeling of its effects in a continuum (right)}. A grain boundary (GB) is a line of disclinations of opposite curvature, pentagonal and heptagonal here, arranged in such a way that the two regions (grains) of the membrane match. The two grains have lattice directions that make an angle $\theta/2$ with respect to the direction the lattice would have in the absence of the GB. Different arrangements of the disclinations, always carrying zero total curvature, correspond to different $\theta$s, the allowed number of which is of course finite, and related to the discrete symmetries of the lattice (hexagonal here). Other arrangements are shown in this paper, see Fig.\ref{fig:GBDoubleColoring}, and more can be found in \cite{hirth1967theory}.  In general, one might expect that the angle of the left grain differs in magnitude from the angle of the right grain, $|\theta_L| \neq |\theta_R|$, nonetheless, high asymmetries are not common, and the symmetric situation depicted here is the one the system tends to on annealing \cite{C5NR04960A}. Therefore, we use the picture here as the prototypical GB, where grain A and grain B are related via a parity ($x \to - x$) transformation. With this, the right-handed frame in grain A is mapped to the left-handed frame in grain B, so that the net effect of a GB is that two orientations coexist on the membrane, and a discontinuous change happens at the boundary. If one wants to trade this discontinuous change for a continuous one, an equivalent coexistence is at work in the non-orientable Moebius strip. One way to quantify the effects of different $\theta$s, as explained in the main text, is to relate a varying $\theta$ to a varying radius $R(\theta)$ of the Moebius. Notice that the third spatial axis is an abstract coordinate, $\tilde{z}$, whose relation with the real $z$ of the embedding space is not specified.}
  \label{fig:GrainBoundaryTwoChiralities}
\end{figure}

The explicit relation between Burger vectors and torsion can be written as \cite{Katanaev2005}
\begin{equation}\label{torsion-Burgers}
b^{i}=\iint_{\Sigma}T^{i}_{\mu\nu}dx^{\mu}\wedge dx^{\nu}\;,
\end{equation}
where the surface $\Sigma$ has a boundary enclosing the defect. This means that the torsion tensor is the surface density of the Burgers vector.

It is only when torsion is zero that the two fields, $\omega$ and $e$, are not independent and, usually, the theory is formulated in terms of $e^a_\mu$, that is $\omega (e)$. This is the case of standard general relativity\footnote{It is also the case of the gravitational Chern-Simons gravity \cite{Deser:1981wh,Deser:1982vy}, also known as conformal gravity \cite{Horne:1988jf} (see also \cite{Guralnik:2003we}).}. For $(2+1)$-dimensional pure gravity, i.e., without matter fields, the gauge transformations of $e^a_\mu$ are on-shell diffeomorphisms only when torsion is zero, see \cite{Witten1988}. Therefore, when torsion is nonzero two important things happen. First, the spin connection acquires an extra, nonmetric, contribution $\tensor{\omega}{^{a}_{\mu}} = \tensor{\widetilde{\omega}}{^{a}_{\mu}}(e) + \tensor{\kappa}{^{a}_{\mu}}$ ($\tensor{\widetilde{\omega}}{^{a}_{\mu}}$ is the \emph{torsionless spin connection} and depends only on the vielbein $e$, while $\tensor{\kappa}{^{a}_{\mu}}$ is called \emph{contorsion tensor}); second, the vielbein is no longer the gauge field associated to local translations, even in the pure gravity case.

The above is well known, not only in condensed matter, but even in the context of the elastic theory formulation of gravity \cite{Kleinert:1989ky,Katanaev:1992kh}. We want to introduce a less known defect, that plays a mayor role in this paper (and in many condensed matter studies), that is the \textit{grain boundary} (GB) \cite{hirth1967theory}. This is a boundary between two regions (grains) that have different relative orientation, given by the so-called misorientation angle $\theta$. Given the hexagonal structure, misorientation angles are constrained to be only certain specific values, the most common (stable) being $\theta = 21.8^o$, and $\theta = 32.3^o$, see, e.g., \cite{Yazyev2014, Yazyev2010}. See Fig.\ref{fig:GrainBoundaryTwoChiralities}. There exists \cite{Yazyev2010,hirth1967theory} a relation (the Frank formula) between $\theta$ and the resultant Burger vector, obtained by adding all Burger vectors $\vec{b}$s cut by rotating a vector $V$, laying on the GB, of an angle $\theta$ with respect to the reference crystal (horizontal lines in Fig.\ref{fig:GrainBoundaryTwoChiralities}).

In the following Section we put forward our conjectures on how to describe, in a field theoretical language, GBs and related scenarios where the two Dirac points coexist. In Section \ref{sec:Witten} we analyze the effects of these conjectures on the emergence of a gauge/gravity field theory description, while in Section \ref{sec:USUSY} we probe the internal symmetry view-point. Finally, in Section \ref{sec:conc}, we draw our conclusions. Some details of the discussion are left to three Appendices.

\section{Grain boundaries and the Moebius strip correspondence}
\label{sec:condmatt}

\begin{figure}%
    \centering
    \includegraphics[width= .6\textwidth, angle = 0]{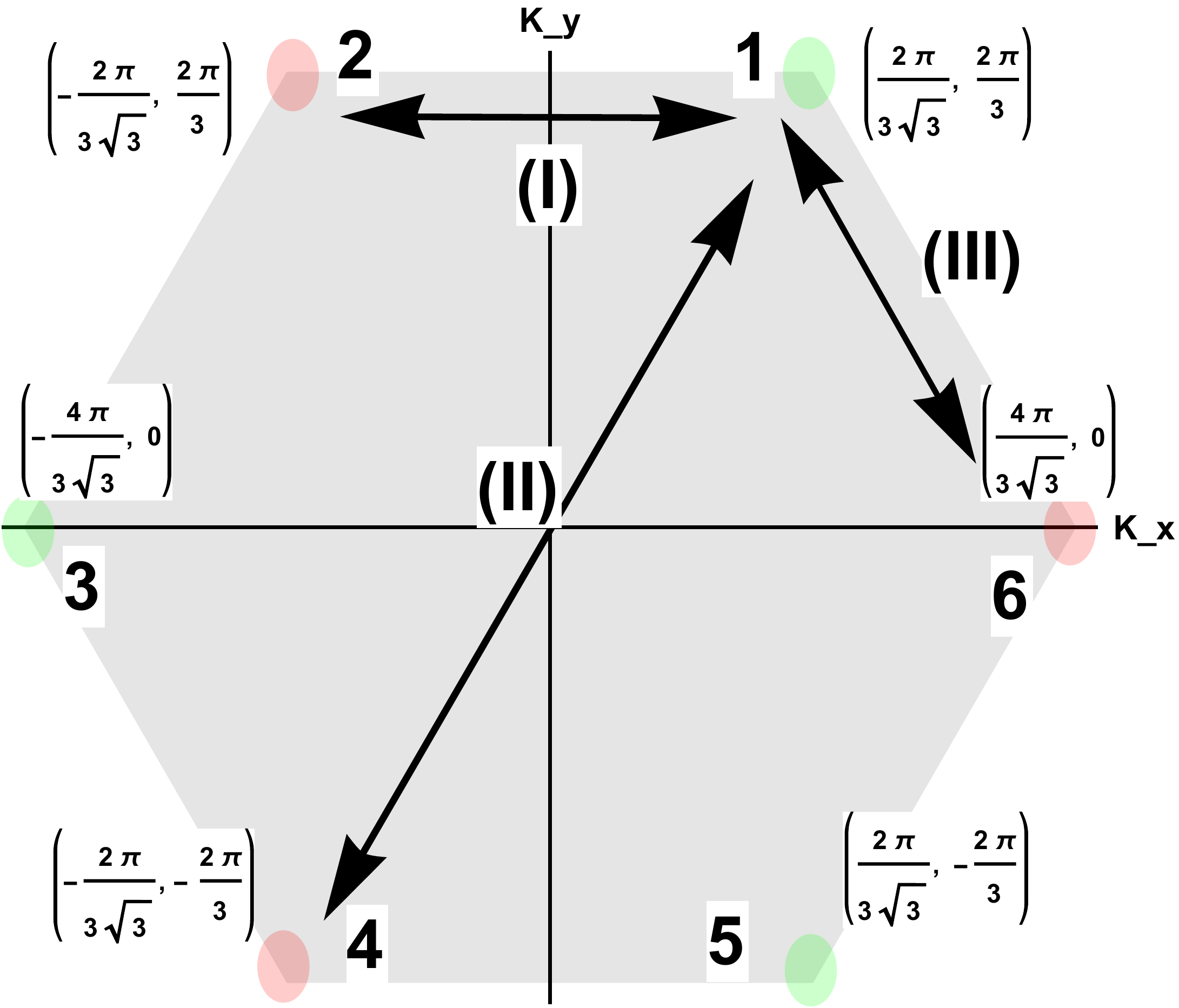}
    \caption{{\bf Our choice of first Brillouin zone and Dirac points.} Points with the same color are connected through vectors of the reciprocal lattice, hence are equivalent. There are only two non equivalent families, and we are free to pick any member of the family as representative. The physics does not depend on this choice, but the description in terms of Dirac Hamiltonians and fields, does depend on that. Certain choices are such that the transformation relating the two points can be cleanly read as the same transformation relating the two Dirac Hamiltonians living in the neighbor of such points. This is explained in the main text and in \ref{appendix_Dirac_points}, where these special cases are called ``suitable''. As an example, in the figure we indicate how the Dirac point $1$ is related to each of the non equivalent points: $x$-parity for case (I), $1 \leftrightarrow 2$; full inversion for case (II), $1 \leftrightarrow 4$; and $\pi/3$ rotation for case (III), $1 \leftrightarrow 6$.}%
    \label{fig:HexagonBrilloin}%
\end{figure}

Although we define the idealized GB as a region only characterized by the misorientation angle $\theta$, once we deal with real materials, we know the interface line which divides the two domains of the honeycomb is finite. Applying the Frank formula, we can relate the distance of the interface line $D$, the misorientation angle $\theta$ and the resultant Burger vector $\vec{B}$. Therefore, for such a finite domains of GBs,  through (\ref{torsion-Burgers}) we know there is a nonzero torsion associated to them\footnote{The reverse is not true. Indeed, we have torsion even when a GB cannot be defined, like the case of a well-defined single $\vec{b}$ of Fig.\ref{fig:edge-dislocation}.}. However, from now on, we will concentrate only on the effects of the misorientation angle $\theta$ on the continuum limit (considering an idealized infinite line interface GB), living aside for the moment the explicit role played by this torsion, on which we shall comment at the end of the paper. For more details on these points see \ref{appendix_torsionGB}.

We make now a threefold proposal, to model the continuum membrane/spacetime associated to the existence of GBs in the lattice. The Dirac field theory, emerging in this approximation, will live on such spacetime.

First, we interpret the different orientations on the two sides of the boundary, as the effect of a reflection of the reference frame ($x \to - x$), that is a \textit{parity} transformation. This interpretation is justified as illustrated in Fig.\ref{fig:GrainBoundaryTwoChiralities}: if in one grain one defines a positive angle as the smallest angle between the grain direction and the direction of the reference crystal, then (if necessary, through a redefinition that makes the alignment angle  \cite{C5NR04960A} vanish \cite{hirth1967theory}) the angle in the other grain is clearly obtained through $x \to - x$. With reference to Fig.\ref{fig:GrainBoundaryTwoChiralities}, in the grain A the natural frame is left handed, while in the grain B the frame is right handed.

\begin{figure}%
    \centering
    \subfloat{{\includegraphics[width= .8\textwidth]{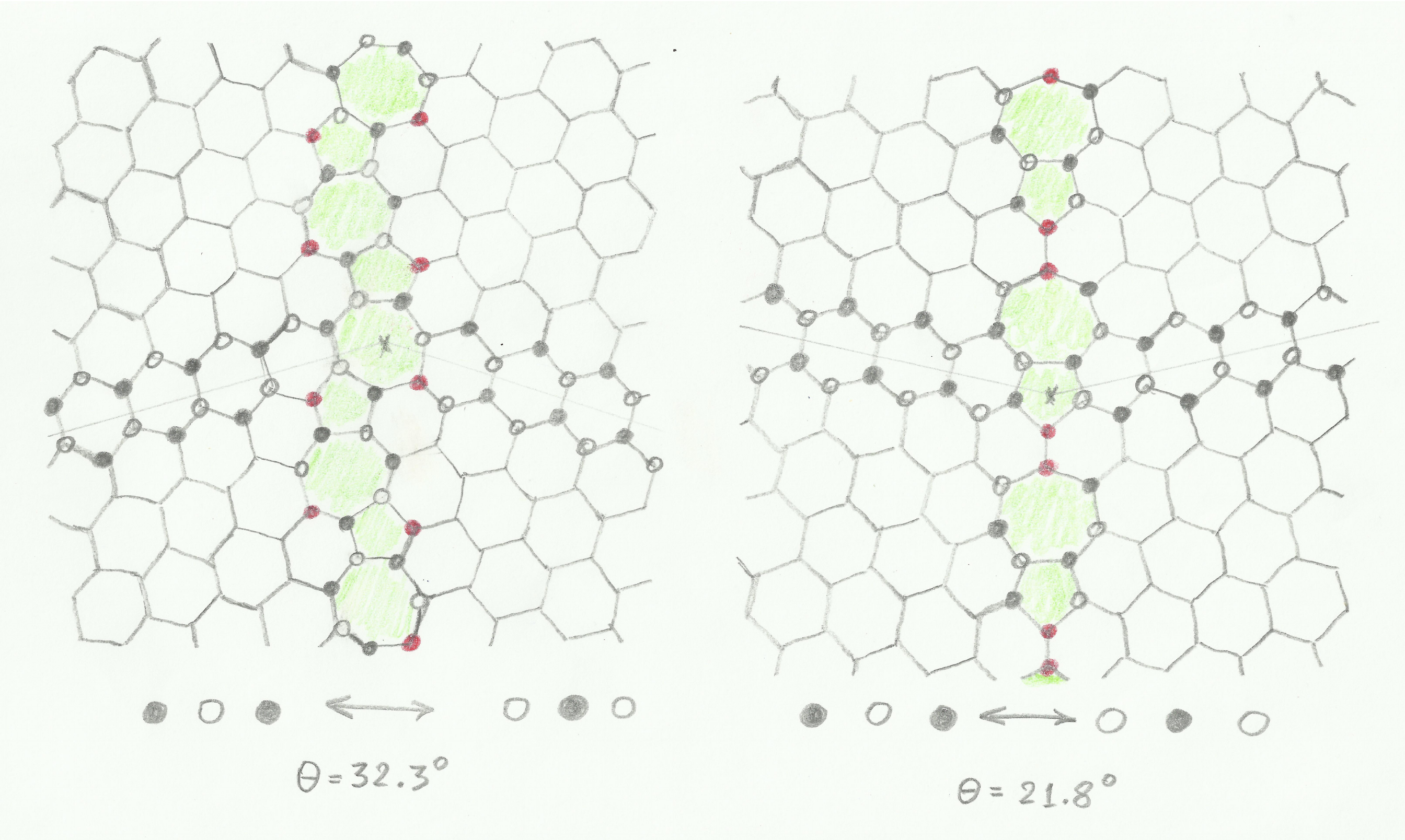} }}%
    \qquad
    \subfloat{{\includegraphics[width= .8\textwidth]{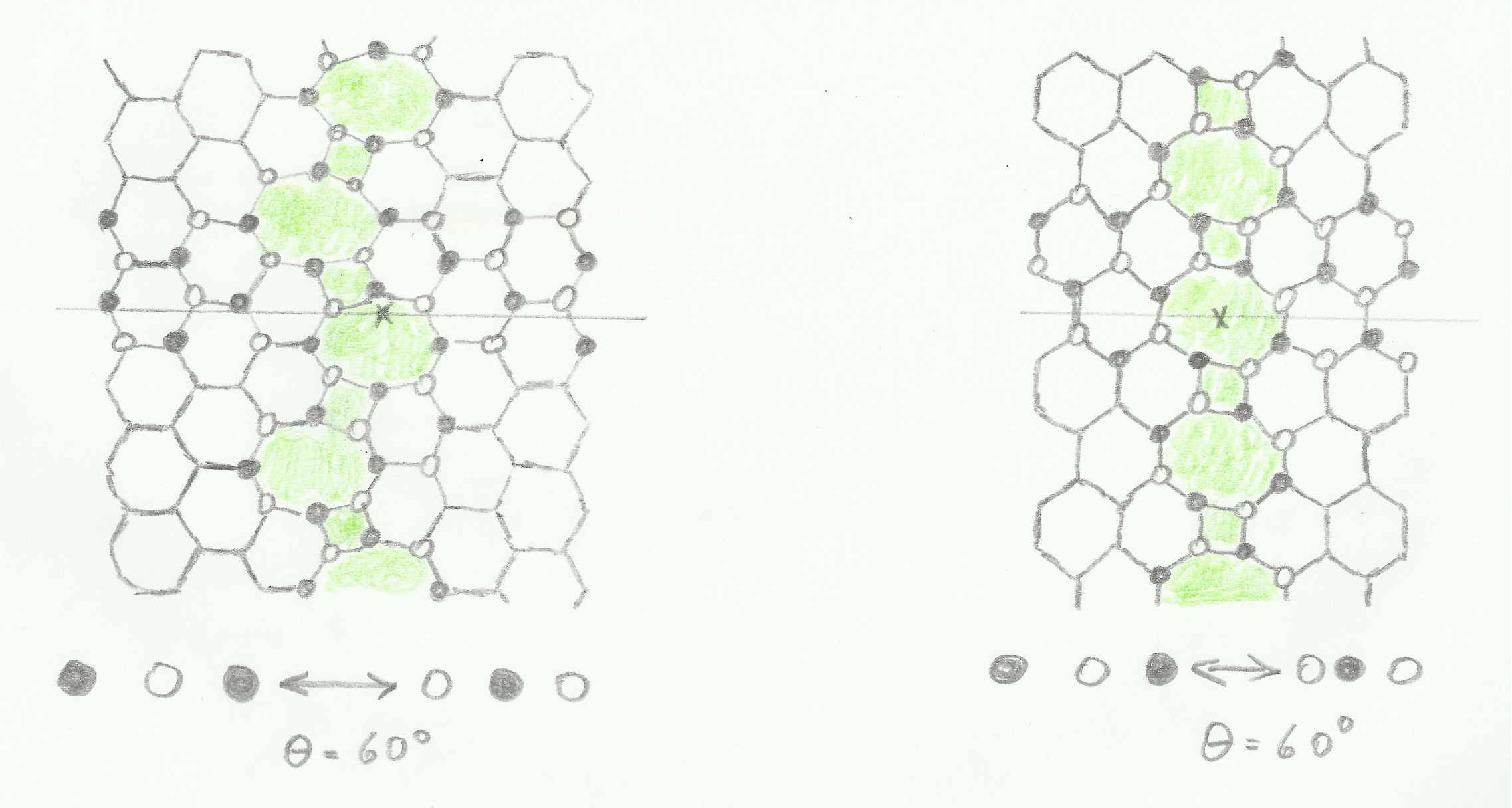} }}%
    \caption{{\bf Four examples of realistic GBs for hexagonal lattices.} The first two GBs (upper row) are very common in graphene \cite{Yazyev2014} and correspond to $\theta = 32.3^o$ (left) and $\theta = 21.8^o$ (right). The last two GBs (lower row) are less common structures (because, the squares and octagons, due to their high curvature, are energetically disfavored disclinations), but still theoretically possible. Both of them correspond to $\theta = \pi/3 = 60^o$, that is a symmetry angle for the lattice. We can introduce a ``coloring pattern'', that is: nearest neighbors to a point of a given color have the other color. For graphene the colors are two, related to the two sublattices, and in turn with the valley degree of freedom. In all four cases, in general, the coloring pattern gets reversed in going from one grain to the other. For the sake of clarity, we only picture this for a couple of lines of hexagons on both sides of the boundaries, always pictured in green. In the case of the first raw, this also comes along with the appearance of lattice nodes that necessarily carry both colors (frustration) and are indicated in red. In the case of $\theta = 60^o$, far from the boundary, the change of coloring pattern is the only way to distinguish this case from the case of no boundary at all, say it $\theta = 0$. Therefore, in general, both valleys (Dirac points) are necessary to describe these systems.}%
    \label{fig:GBDoubleColoring}%
\end{figure}

Second, in the spirit of the continuum theory we are seeking, where the Dirac field theory lives, we replace the laboratory situation, with a correspondent\footnote{In the correspondence, some aspects of the original physics are recovered, some are lost. For a discussion of a general kind, but closer to these scenarios, see \cite{Iorio:2015iha}, while, for a more philosophical one see \cite{Dardashti2016arXiv160405932D}.} situation for which the discrete parity reflection, that happens discontinuously at the boundary, is traded for a continuous transformation, namely a full rotation around an axis living in the third dimension of a corresponding twisted space. The geometry we are evoking, then, is clearly that of the Moebius strip, that is a nontrivial fiber bundle \cite{Nakahara} with base space $S^1$, and structure group $G = \mathbb{Z}_{2}$, where $S^1$ is the circle of radius $R(\theta)$, see Fig.\ref{fig:GrainBoundaryTwoChiralities}.

Third, we choose to model this situation with a varying $R(\theta)$, whose limits are $\lim_{\theta \to 0} R = \infty$ and $\lim_{\theta \to \pi/3} R = R_{min}$. This way, the effects of a nonzero misorientation angle are gone when the radius is infinite because the path to take a full turn, $L = 2 \pi R$, is infinite (the vector never comes back, and the rotation around the third axis is indeed a translation). When the $R = \infty$ limit is reached, we also suppose that the bundle trivializes to a cylinder, $S^1 \times I$.

Another option for modeling this latter feature is to fix $R$, hence $L$, for all $\theta$s, and require that the structure group $G$ acts repeatedly, and periodically in $\theta$, on the base space, $S^1$. This way of modeling would correspond to have $m$ twists after a full turn, corresponding to $m$ actions of the elements of $G$. On the other hand, when $m = 2k+1$ this is topologically equivalent to one (1) twist, and when $m = 2k$ this is like no twists (0), so the group is always $(0,1)$, that is $\mathbb{Z}_{2} = G$. Since we are in a corresponding space, we do not take into account strain and other deformations that would, of course, make $m$ twists physically distinguishable from  any $m'$ twists. Therefore, we prefer to stick to the previous modeling, that is, one GB/one twist, and leave the correspondence with multiple-twist to a multiple-GB scenario.

So far all considerations stemmed from the membrane alone, and from the transformations of a Cartesian coordinate frame (passive view), or of a \textit{vector} (active view) on the membrane. On the other hand, graphene is a scenario where it is the structure of the space responsible for the emergence of specific field structures, that is the \textit{spinor}. In turn, the structure of the space is also related to the structure of the reciprocal $k$-space, in particular with the existence of the two Dirac points that are at the core of the interest of this paper.

\begin{figure}
  \centering
  \includegraphics[width=1\textwidth ]{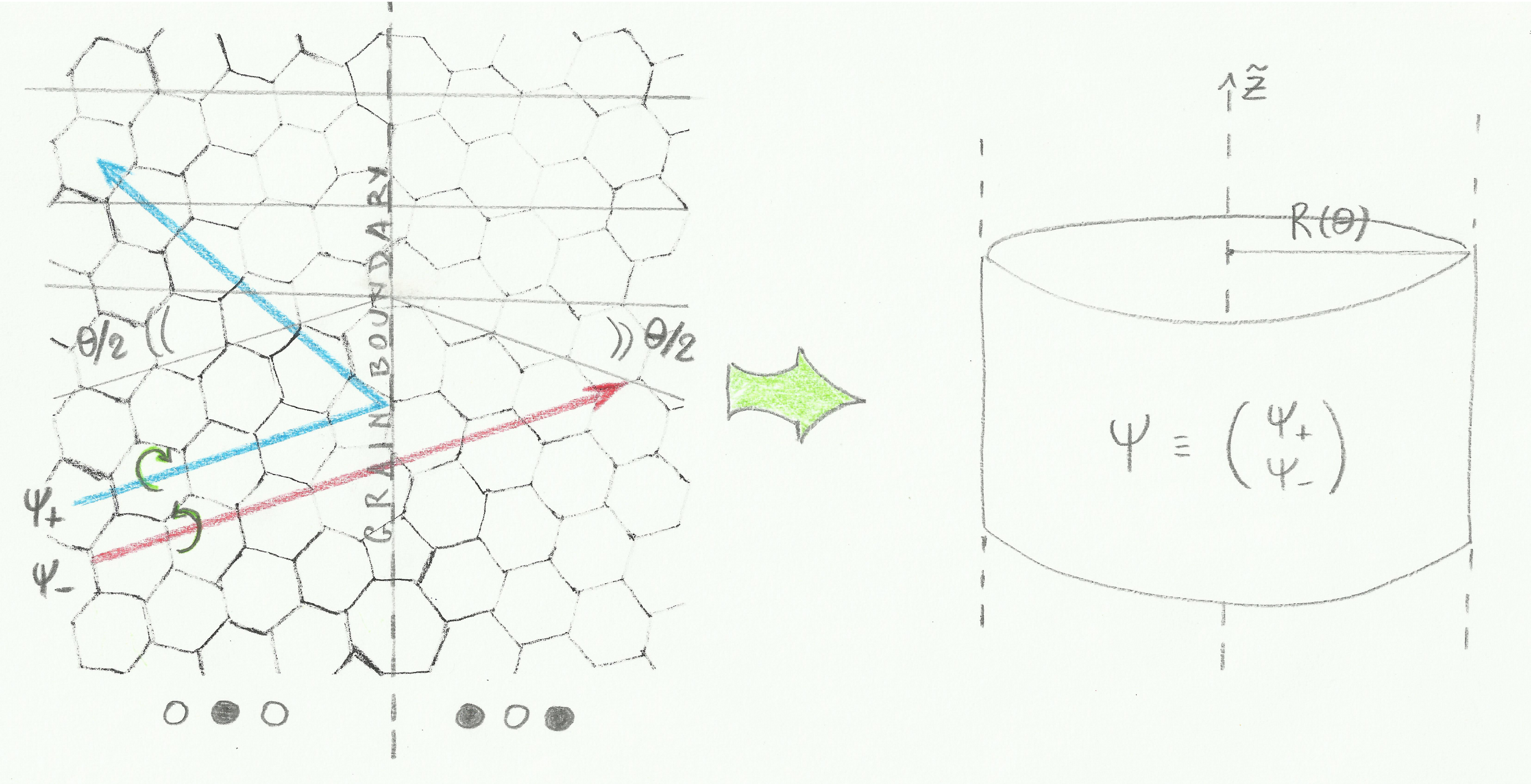}
  \caption{{\bf A possible modeling of the effects of a GB in a continuum description.} The coexistence on the membrane of two parities is associated with the coexistence of the two coloring patterns, hence with the coexistence of the two spinors, $\psi_\pm$, each emerging at a different Dirac point, each with a specific parity.  As before, see Fig. \ref{fig:GrainBoundaryTwoChiralities}, the discontinuous change happening at the GB (left) is traded for a Moebius-like continuous structure (right), but this time the action of the $\mathbb{Z}_{2}$, that makes the Moebius a non trivial bundle, is passed on to the spinor fields living on a cylinder (whose third spatial axis is the abstract $\tilde{z}$, not to be identified with the actual $z$ of the embedding space) but related by a $\mathbb{Z}_{2}$ parity symmetry, $\psi_+ \leftrightarrow \psi_-$. This makes a four component Dirac spinor, $\Psi = (\psi_+, \psi_-)^T$. As before, within the range $\theta \in (0, \pi/3)$), the larger the $\theta$, the smaller $R(\theta)$, the radius of the $S^1$. In the figure on the left we also picture the so called ``valley filtering effect'' \cite{Yazyev2014}, for which a spinor with a given associated coloring pattern selects the grain with the same coloring pattern. This we take as a phenomenological indication of how this property of the membrane and a property of the field get intertwined in these models, where all participating fields (either geometric/spatiotemporal or matter fields), emerge from a single underlying structure.}
  \label{fig:Moebius}
\end{figure}

There are two Dirac-like Hamiltonians, one for each non-equivalent Dirac point. For some particular descriptions, the two Dirac points are related by a change in the sublattice, $a \leftrightarrow b$, that means a change in the coloring pattern of the lattice: $ \circ \bullet \circ \leftrightarrow \bullet \circ \bullet$. Although this only happens for certain suitable descriptions (e.g., even for the same Dirac points we have different Hamiltonian choices, see \ref{appendix_Dirac_points}), since the actual condensed matter physics is independent from this choice, we can always pick up a description where this is true, see Fig. \ref{fig:HexagonBrilloin}.

In the language of condensed matter, this is the so-called ``valley degree of freedom'' \cite{PacoReview2009}, denoted here by $\lambda = \pm$. As before, not all choices of the pair of Dirac points have a direct interpretation in terms of parity transformations, but, since the physics must be independent from such choice, we may as well choose the pair that is most suitable for a clean parity transformation in the Dirac language. In \ref{appendix_Dirac_points} we explain all of that in detail. What we need here of that discussion is the following statement: When there are phenomena with a change of coloring pattern/valley, like for  GBs (as we show in the Fig.\ref{fig:GBDoubleColoring}), we need both Hamiltonians, and the change of parity is realized by $\psi_+ \leftrightarrow \psi_-$.

Thus, the key role of the $k$-space in this context is to furnish an extra property of these membranes, related to the parity of the Dirac field leaving on it, $\chi = \pm$. Notice the relation $\lambda = \xi \chi$, where $\xi = +$ for a particle, and $\xi = -$, for a hole/antiparticle \cite{RevModPhys.83.1193}, that tells us that particles (and antiparticles) at inequivalent Dirac points, have opposite parity. With the above in mind, in the field theory correspondence we are building-up, this motivates the reinterpretation of the Moebius geometry, a membrane where two orientations coexist, in terms of a membrane where two parities coexist, hence the two spinors are necessary at once, $\Psi = (\psi_+, \psi_-)^T$. This is illustrated in Fig.\ref{fig:Moebius}.

\begin{figure}[H]
    \centering
     \includegraphics[width= .8\textwidth]{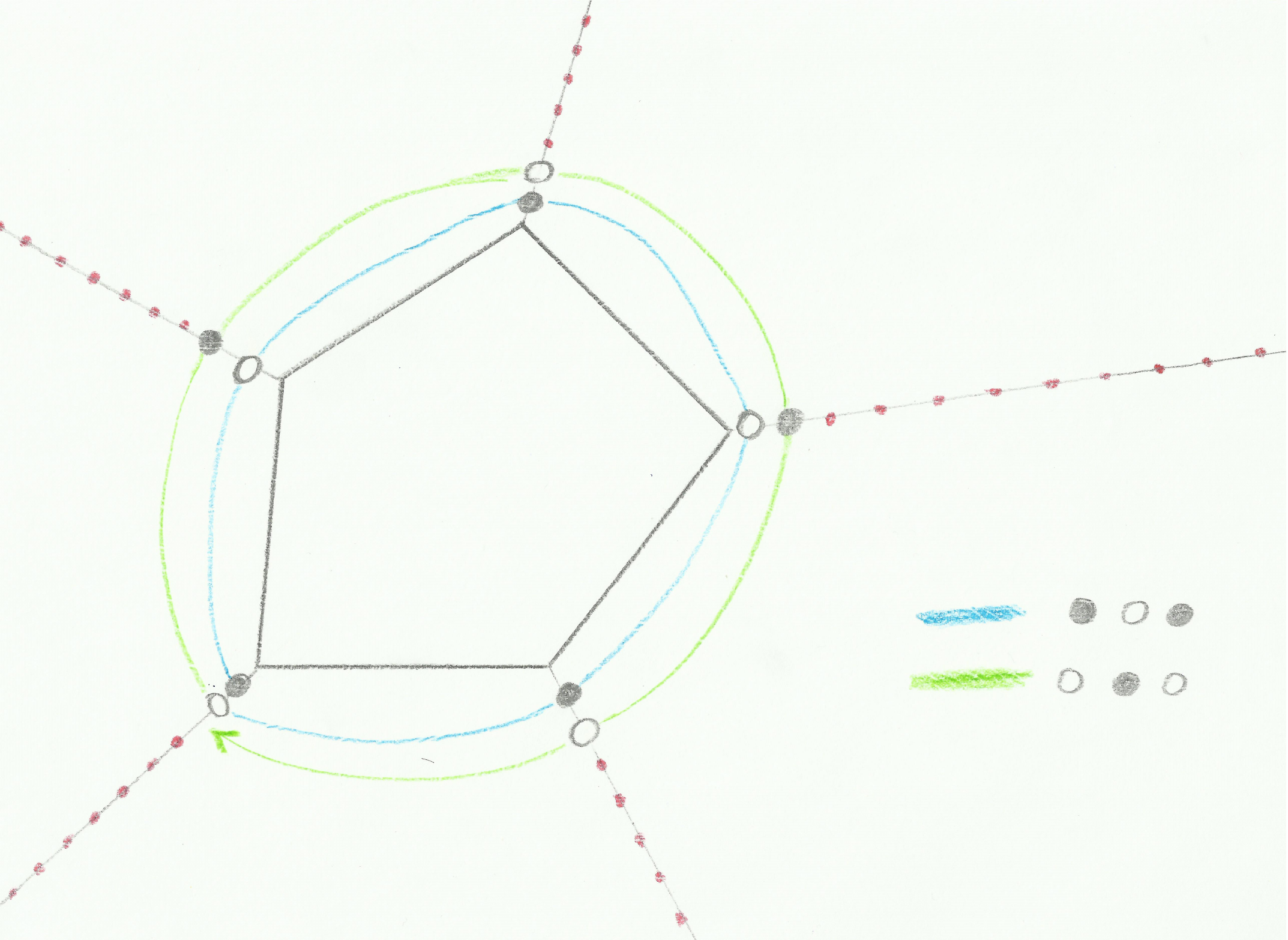}
    \caption{{\bf Pentagonal defect and its generalized grain boundaries}. Having an odd-sided disclination defect implies, by itself, the necessity for the coexistence of the two coloring patterns, all over the membrane. Indeed, if we start a tour of the defect from a given point (here: the black circle of the lowest vertex, where the blue line starts) and take a full turn (going clockwise here), that same point necessarily changes its coloring, and at a second turn (the green line here), the coloring pattern has changed. Therefore, from that point it starts a frustration/double-coloring line (indicated with red spots here) that separates two ``grains'', differing \textit{only} for coloring pattern, hence possible to interpret as grains with a misorientation angle equal to a symmetry of the lattice, $\theta = \pi/3$ (see also Fig. \ref{fig:GBDoubleColoring}). This red line we might identify with as a GGB. A field sensitive to the coloring pattern will spot the necessity for two such patterns. Of course, since there is no real line of defects, one can start to turn from any other vertex, hence here the GGBs are as many as the (odd) number of vertices of the defect, as indicated. Another important difference with respect to a standard GB is that the change of coloring patterns, in this case, comes through \textit{circling} around a tip (conical geometry), rather than \textit{translating} through a line (cylindrical geometry).}%
    \label{fig:PentagonDoubleColoring}%
\end{figure}

Let us close this part by proposing a unifying view. We define as \textit{generalized grain boundary} (GGB) a standard GB or a boundary between grains differing only in the coloring patterns ($\circ \bullet \circ \leftrightarrow \bullet \circ \bullet$), not in the orientation, see Fig.\ref{fig:GBDoubleColoring}.

When we have a single disclination defect (say a pentagon), one possible interpretation is that the misorientation angle, $\theta$, reached a value that is a symmetry angle of the lattice ($\theta = \pi / 3$, for $n=5$). This way, a line stemming from a vertex can be seen as GGB, see Fig.\ref{fig:PentagonDoubleColoring}. Of course, the situation with a GGB (single defect) is different from a real GB (line of defects), because to the former it corresponds the geometry of a cone, while to the latter, through our modeling, it corresponds a Moebius strip/cylinder geometry. Nonetheless, the similarities just described allow for two choices that would include the disclinations in the general discussion.  The first, is to use the Moebius/cylinder model for all angles, including the lattice-symmetry values, $\theta = n \pi / 3$, for which, in fact, the best model is the cone. The second, is to use the cone model even for angles $\theta \neq n \pi/3$, corresponding to disclinations in a hexagonal lattice, that is, one could think of cones with apex angle varying continuously from small to large values\footnote{In a one-defect-per-tip situation, there are only $4$ possible angles, but in a many-defect-per-tip situation, there can be many more}. In both cases, though, one would need another parameter besides $\theta$ to characterize the defect, that is, e.g., the height of the cone, $h$, so that $R(\theta) \to R(\theta,h)$.

\begin{figure}
  \centering
  \includegraphics[width=1\textwidth]{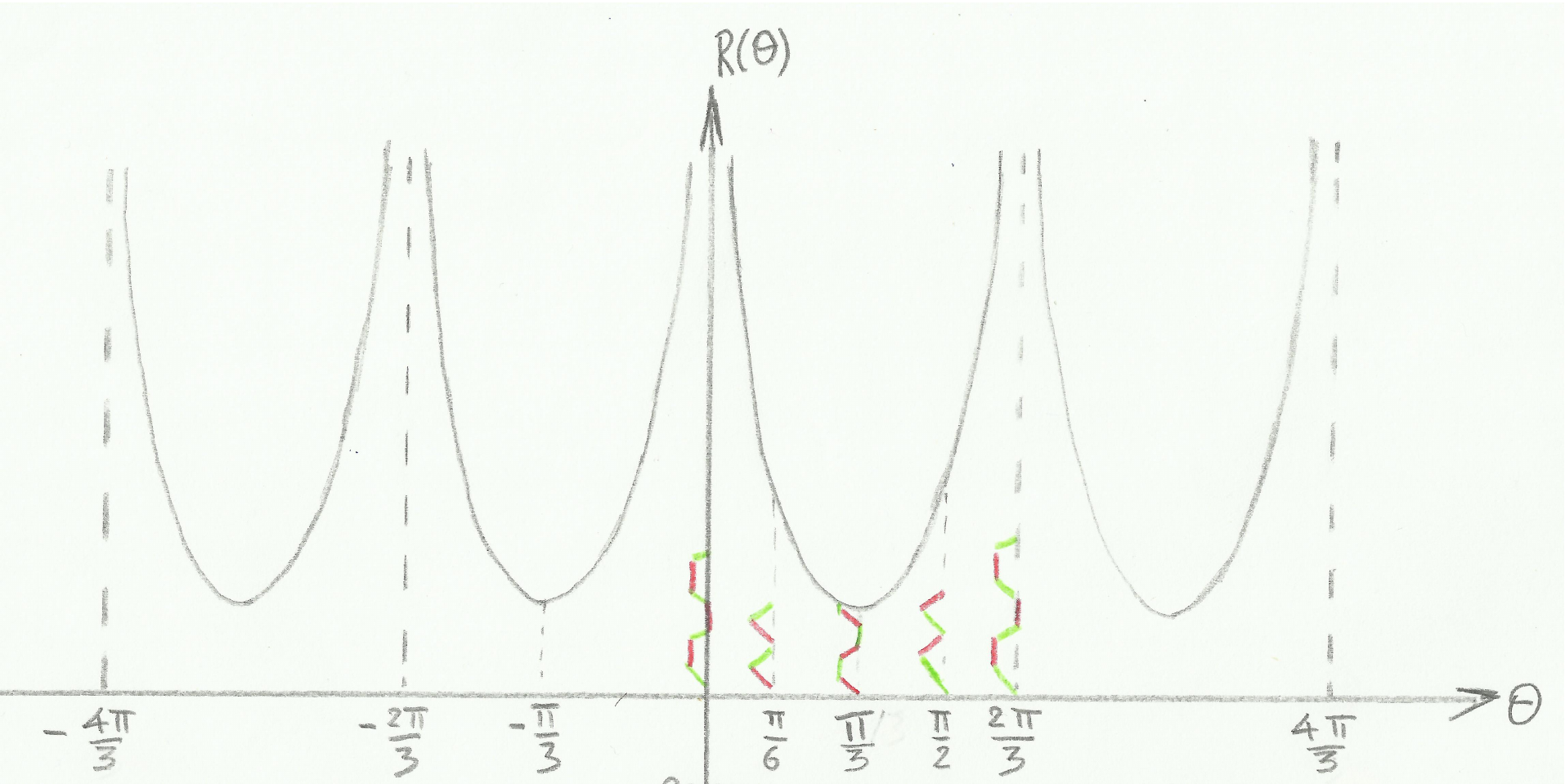}
  \caption{{\bf The posited periodic behavior of $R(\theta)$.} The effects of a nonzero $\theta$, in our picture, become stronger in the range $\theta \in (0, \pi/3)$. Indeed, $\theta = 0$ corresponds to the undeformed lattice (e.g., of the armchair type in this figure, see drawing on the vertical axis), to which we make to correspond $R = \infty$, for which the translation along the $S^1$ never comes back, hence stays a translation. On the other end, $\theta = \pi/3$, the misorientation angle equals a symmetry angle of the lattice, for which the initial lattice orientation is recovered, but with the opposite coloring pattern (here indicated with the armchair of the opposite coloring pattern with respect to the initial armchair). To this we make to correspond the minimal value of $R$, i.e., the maximal effect. Notice also that at $\theta = \pi/6$ the armchair turns into a zig-zag.
  In the next interval, $\theta \in (\pi/3, 2 \pi/3)$, the shape of $R(\theta)$ is necessarily symmetric, because grains rotated of $\alpha$ are indistinguishable from grains rotated of $\pi/3 + \alpha$, for $0 < \alpha < \pi/3$, if were not for the coloring patterns, and, although we do add a label in the plot taking that into account, in our modeling of the effect of GBs the coloring pattern degree of freedom goes to the ``matter fields $\psi$'', not to the membrane. Thus, when the GB is there, hence, in general, the double coloring pattern is necessary (which makes a clear distinction between $\theta = 0$ and $\theta = \pi/3$), if nothing else is considered (like, e.g., strain, or the energy cost for the formation of the defect) in the model we cannot know whether the effect is due to $\alpha$ or to $\pi/3 + \alpha$, or to the negative values (corresponding to opposite angle orientation, e.g., in a GGB picture, to negative curvature defects like the heptagon). This process goes until the next symmetry angle of the lattice, $2 \pi / 3$, corresponding also to a restoration of the original coloring pattern. This figure is an idealization for three reasons: i) $\theta$ is continuous; ii) there is no difference between $\theta = 0$ and $\theta = 2 \pi /3$; iii) the periodic plot is repeated over a large range of $\theta$. }
  \label{fig:RofTheta}
\end{figure}

For the considerations of this paper, we shall use the first interpretation, and we shall refer to a radius/characteristc length $R(\theta)$, whose phenomenological behavior is depicted in Fig.\ref{fig:RofTheta}.

\section{High-energy-theory correspondence I: Spatiotemporal approach and Witten three-dimensional gauge-gravity theory}
\label{sec:Witten}

It is an old idea to see gravity as a gauge theory of the Poincar\'{e} group. The connection for such group is
\begin{equation}\label{gaugegravity}
   A_\mu =  \frac{i}{2} \tensor{\omega}{^{a b}_{\mu}} \mathbb{J}_{a b}  +  i \tensor{e}{^{a}_{\mu}} \mathbb{P}_{a}  \;,
\end{equation}
with the spin connection, $\omega^{a b}_\mu$, as the gauge fields for (pseudo) rotations of the local Lorentz transformations, and the vielbein, $\tensor{e}{^{a}_{\mu}}$, as the gauge fields for translations/diffeomorphisms. However, as is well known, this programme works properly only in three dimensions without matter fields \cite{Witten1988}, that is the case of graphene spacetimes without fermions on them. For a nice account of this story, see, e.g., \cite{Z-book} and references therein.

With the discussion of the previous Section in mind, we want to pursue this programme by studying a minimal coupling of $ A_\mu$ in (\ref{gaugegravity}) with the Dirac fermions of graphene. Our logic here is that, when a GB is present, three things happen: i) torsion is present on the membrane; ii) a characteristic length, $R(\theta)$, appears; iii) both Dirac points, each with its own valley number, are at work, hence it is better to introduce the four-spinor $\Psi =\left(
                                             \begin{array}{c}
                                               \psi_{+} \\
                                               \psi_{-} \\
                                             \end{array}
                                           \right)$.

Therefore, as the most natural candidate for the gauge field of spatiotemporal nature, emerging from this picture, we propose the gauge field associated to the following algebra\footnote{As we are in $(2+1)$ dimensions, we can use the dual index notation $\omega^{a}=\tensor{\epsilon}{^{a}_{bc}}\omega^{bc}$ and $\mathbb{J}_{a}=\tensor{\epsilon}{_{a}^{bc}}\mathbb{J}_{bc}$, with convention $\epsilon_{0123}=1$.}
\begin{eqnarray}\label{AdS_algebra}
\left[\mathbb{J}_{a},\mathbb{J}_{b}\right]&=&i  \, \tensor{\epsilon}{_{ab}^{c}} \, \mathbb{J}_{c} \;, \\
\left[\mathbb{J}_{a},\mathbb{P}_{b}\right]&=&i  \, \tensor{\epsilon}{_{ab}^{c}} \, \mathbb{P}_{c} \;, \\
\left[\mathbb{P}_{a},\mathbb{P}_{b}\right]&=&i  \, \lambda(R)  \,  \tensor{\epsilon}{_{ab}^{c}} \,  \mathbb{J}_{c} \;,
\end{eqnarray}
where, all generators are $4\times4$ matrices, and, the Lorentz generators are {\it diagonal in the valley degree of freedom}
\begin{equation}\label{FirstJ}
 \mathbb{J}_{a}=\left(
                                             \begin{array}{cc}
                                               J^{+}_{a} & 0 \\
                                               0 & J^{-}_{a} \\
                                             \end{array}
                                           \right) \,,
\end{equation}
whereas the translation generators are {\it off-diagonal in the valley degree of freedom}
\begin{equation}\label{FirstP}
\mathbb{P}_{a}=\left(
                                             \begin{array}{cc}
                                               0 & P^{+}_{a} \\
                                               P^{-}_{a} & 0 \\
                                             \end{array}
                                           \right) \,.
\end{equation}
An explicit representation of the generators $\mathbb{J}_{a}$ and $\mathbb{P}_{a}$ of the form \eqref{FirstJ} and \eqref{FirstP} is given in \ref{appendix_J}.

The $\lambda$ parameter has dimension of $[\mbox{length}]^{-2}$, and we take it to be
\begin{equation}\label{lambdaR}
  \lambda \equiv 1/R^2(\theta) \,.
\end{equation}
This means, from the graphene point of view, that when $\theta \neq 0$ (that is, when $R \neq \infty$), the $\mathbb{P}$ generators set in and mix the valleys, while the $\mathbb{J}$ generators keep acting independently in each valley. From the fundamental/high-energy point of view, this means to extend the Lorentz group, $SO(2,1)$, to de Sitter, $SO(3,1)$ or Anti de Sitter, $SO(2,2)$, group, when $\lambda \neq 0$. One key point of the latter extension is that translations become, in fact, (pseudo) rotations, e.g., for de Sitter, in a sphere of radius $R$. In our view, this rotations are around an axis living in a third dimension, as the analysis below will show.

Before moving on, let us make a comment here. (Anti-)de Sitter space has (negative) positive \emph{constant} curvature. Strictly speaking, this does not match our picture of the previous Section, where we have a Moebius/cylinder which is characterized by the radius $R$ only in \emph{one} of the directions, perpendicular to the GB, but does not in the other directions, as (Anti-)de Sitter space requires. Therefore, the Moebius/cylinder approach of the previous Section points out to a flat object from a Riemannian point of view. Nonetheless, we have to think here of each GB as a basic building block to construct the given manifold/spacetime, whose precise structure depends upon the details\footnote{For instance, six such boundaries, each joining two vertices of the 12 necessary to close an icosahedron, are at work in one possible modeling of the giant fullerene molecules \cite{GONZALEZ1993771}.}. The overall effect of this patchwork is then to have a (average) characteristic radius $R$ in all directions, although, as it will be clearer later, in an abstract space where at least one of the dimensions (the third spatial dimension $\tilde{z}$) is purely abstract.

We are ready now to move, with this theoretical machinery, to the condensed-matter-inspired scenarios of graphene. To summarize, our goal will be to see whether the covariant derivative
\begin{equation}
\overrightarrow{D}_{\mu}\Psi=\partial_{\mu}\Psi + \omega^{a}_{\mu}\mathbb{J}_{a} \Psi + i e^{a}_{\mu}\mathbb{P}_{a} \Psi \;,
\end{equation}
makes sense in this context.

\subsection{Lorentz spin-$1/2$ representation in $(2+1)$ dimensions and its extensions}
\label{sec:Lorentz_extension}

To keep the discussion as general as possible, we denote the generators of this extended Lorentz algebra, either for the de Sitter or anti-de Sitter case, as $\mathbb{J}_{AB}$, were $A,B\in\{0,1,2,3\}$ and $a,b\in\{0,1,2\}$, keeping in mind that we are in $(2+1)$ dimensions, such that $\mathbb{J}_{AB}$ are the $SO(2,1)$ Lorentz generators for $A,B\in\{0,1,2\}$ and $\mathbb{P}_{a}=\mathbb{J}_{A3}$ are the \emph{extra} generators, which could represent either the de Sitter or anti-de Sitter algebra (or the Poincar\'e algebra after a In\"on\"u-Wigner contraction \cite{Gilmore-book}).

The generators $\mathbb{J}_{AB}$ must fulfill the (A)dS algebra
\begin{equation}\label{AdS_algebra}
\left[\mathbb{J}_{AB},\mathbb{J}_{CD}\right]=i\left(\eta_{AD}\mathbb{J}_{BC}-\eta_{AC}\mathbb{J}_{BD}+\eta_{BC}\mathbb{J}_{AD}-\eta_{BD}\mathbb{J}_{AC}\right)\;,
\end{equation}
where $\eta_{AB}=\mbox{diag}(+1,-1,-1,\tau)$ being $\tau=+1$ for anti-de Sitter $SO(2,2)$ and $\tau=-1$ for de Sitter $SO(3,1)$ in three dimensions.

We have to find the generators for the (A)dS spacetime group. For that purpose, we take four anticommuting matrices $\gamma^{A}$, satisfying the Clifford algebra
\begin{equation}\label{Clifford_algebra}
\left\{\gamma^{A},\gamma^{B}\right\}=2\eta^{AB}\;.
\end{equation}
As $(\gamma^{A})^{2}=\eta^{AA}\mathbb{I}$, we can pick these matrices such that their Hermiticity condition is
\begin{equation}\label{Hermiticity_gamma}
\gamma^{A\dagger}=\begin{cases}
                    \gamma^{0}\gamma^{A}\gamma^{0}, & \mbox{if } A=0,1,2 \\
                    -\tau\gamma^{0}\gamma^{A}\gamma^{0}, & \mbox{if } A=3\;.
                    \end{cases}
\end{equation}
The direct way to obtain the generators $\mathbb{J}_{AB}$ of the (A)dS algebra is
\begin{equation}\label{def_J}
\mathbb{J}_{AB}=\frac{i}{4}[\gamma_{A},\gamma_{B}]\;,
\end{equation}
because it ensures immediately \eqref{AdS_algebra}. We can convince ourself this is the only way to construct a generator with two antisymmetric Lorentz kind of indexes (see \ref{appendix_J}). Using \eqref{def_J} and \eqref{Hermiticity_gamma} we have the following Hermiticity condition
\begin{equation}\label{Hermiticity_J_extended}
\mathbb{J}^{\dagger}_{AB}=\begin{cases}
                    \gamma^{0}\mathbb{J}_{AB}\gamma^{0}, & \mbox{if } A,B=0,1,2 \\
                    -\tau\gamma^{0}\mathbb{J}_{AB}\gamma^{0}, & \mbox{if } A=3 \mbox{ or } B=3\;.
                    \end{cases}\;.
\end{equation}

An element of this group in the spinorial representation can be written as $g=e^{-\frac{i}{2}\lambda^{AB}\mathbb{J}_{AB}}$, being $\lambda^{AB}=-\lambda^{BA}$ the number coefficients of the transformation. Therefore, a spinor transforms as
\begin{equation}\label{spinor_transformation}
\Psi(x)\to\Psi'(x')=g\Psi(x) \;.
\end{equation}
This means
\begin{equation*}
\overline{\Psi}(x)\to\overline{\Psi}'(x')=\Psi'^{\dagger}(x')g^{\dagger}\gamma^{0}=\Psi^{\dagger}(x)e^{\frac{i}{2}\lambda^{AB*}\mathbb{J}^{\dagger}_{AB}}\gamma^{0}\;.
\end{equation*}
If we want the transformation
\begin{equation}\label{spinor_bar_transformation}
\overline{\Psi}(x)\to\overline{\Psi}'(x')=\overline{\Psi}(x)g^{-1}\;,
\end{equation}
as usual, then $\gamma^{0}g^{-1}\gamma^{0}=g^{\dagger}$. On the other hand, as the Hermitian properties of the generators are \eqref{Hermiticity_J_extended}, the above requirement implies $\lambda^{ab*}=\lambda^{ab}$ ($\lambda^{ab}$ are real numbers) while $\lambda^{a3*}=-\tau\lambda^{a3}$ ($\lambda^{a3}$ are real numbers for de Sitter and imaginary numbers for anti-de Sitter). Note this is strange for the anti-de Sitter case, as we usually take the parameters of the Lie group to be real. Only in this way $\overline{\Psi}\Psi$ is a scalar under (A)dS algebra.

To construct bilinear spinor vectors, we need the following property to hold (see for instance \cite{Peskin})
\begin{equation}\label{gamma_as_vector_prop}
\left[\gamma^{C},\mathbb{J}_{AB}\right]=i\left(\delta^{C}_{A}\gamma_{B}-\delta^{C}_{B}\gamma_{A}\right)\;.
\end{equation}
We can check directly that the generators defined as \eqref{def_J} fulfill this property (see \ref{appendix_J}).
The reason of \eqref{gamma_as_vector_prop} is that we want $g^{-1}\gamma^{A}g={\Lambda}{^{A}}{_{B}}\gamma^{B}$, where ${\Lambda}{^{A}}{_{B}}\simeq\delta^{A}_{B}+{\lambda}{^{A}}{_{B}}$ is the vectorial representation of the (A)dS group. In other words, we want terms like $\overline{\Psi}\gamma^{A}\Psi$ transforming as an (A)dS vector for the index $A$. The problem with anti-de Sitter case is that ${\Lambda}{^{a}}{_{3}}\simeq{\lambda}{^{a}}{_{3}}$ is imaginary, contradicting the fact that ${\Lambda}{^{A}}{_{B}}$ are real matrices (the $SO(2,2 )$ group is the set of \emph{real} matrices preserving the extended metric $\eta_{AB}$ with determinant one). Therefore, the anti-de Sitter case is very pathological.

If we want to allow local (gauge) transformations $g=g(x)$, we need to introduce the spin connection $\Omega_{\mu}=\frac{i}{2}\omega^{AB}_{\mu}\mathbb{J}_{AB}$ transforming as \cite{Nakahara}
\begin{equation*}
\Omega_{\mu}\to\Omega'_{\mu}=g\partial_{\mu}g^{-1}+g\Omega_{\mu}g^{-1}\;,
\end{equation*}
which implies that its components $\omega^{AB}_{\mu}$ transform infinitesimally as
\begin{equation*}
\omega^{AB}_{\mu}\to {\omega '}^{AB}_{\mu}=\omega^{AB}_{\mu}+\partial_{\mu}{\lambda}{^{A}}{_{B}}+{\lambda}{^{A}}{_{C}}\omega^{CB}_{\mu}-{\lambda}{_{C}}{^{B}}\omega^{AC}_{\mu}\;.
\end{equation*}
Now, the covariant derivative is defined as
\begin{eqnarray}\label{covariant_derivative}
\overrightarrow{D}_{\mu}\Psi&=&\partial_{\mu}\Psi+\Omega_{\mu}\Psi=\partial_{\mu}\Psi+\frac{i}{2}\omega^{AB}_{\mu}\mathbb{J}_{AB}\Psi \;,\\ \overline{\Psi}\overleftarrow{D}_{\mu}&=&(\overrightarrow{D}_{\mu}\Psi)^{\dagger}\gamma^{0}=\partial_{\mu}\overline{\Psi}-\frac{i}{2} \overline{\Psi}\mathbb{J}_{AB}\omega^{AB}_{\mu}\;. \nonumber
\end{eqnarray}
Note that $\omega^{ab*}_{\mu}=\omega^{ab}_{\mu}$ and $\omega^{a3*}_{\mu}=-\tau\omega^{a3}_{\mu}$. Definition \eqref{covariant_derivative} ensures the nice property that $\partial_{\mu}(\overline{\Psi}\Psi)=\overline{\Psi}\overleftarrow{D}_{\mu}\Psi+\overline{\Psi}\overrightarrow{D}_{\mu}\Psi$, which could be important in manipulations involving partial integration. More important, definition \eqref{covariant_derivative} also guaranties $\overrightarrow{D}_{\mu}\Psi$ and $\overline{\Psi}\overleftarrow{D}_{\mu}$ transforming as spin-$1/2$ field under local (A)dS group. Indeed, we can check in this case
\begin{eqnarray}\label{covariant_derivative_transf}
(\overrightarrow{D}_{\mu}\Psi)'&=&\partial_{\mu}\Psi'+\Omega_{\mu}\Psi'=\partial_{\mu}g\Psi+g\partial_{\mu}\Psi+g\partial_{\mu}g^{-1}g\Psi+g\Omega_{\mu}g^{-1}g\Psi \\
&=&g\left(\partial_{\mu}\Psi+\Omega_{\mu}\Psi\right)=g\overrightarrow{D}_{\mu}\Psi\;,
\end{eqnarray}
where in the third equality we used the fact that $\partial_{\mu}g^{-1}g=-g^{-1}\partial_{\mu}g$. Analogously, $(\overline{\Psi}\overleftarrow{D}_{\mu})'=\overline{\Psi}\overleftarrow{D}_{\mu}g^{-1}$.

\subsection{$SO(3,1)$ or $SO(2,2)$ invariance in $(2+1)$ dimensions}

To construct a sensible action, we need locally invariant terms under $SO(3,1)$ (de Sitter) or $SO(2,2)$ (anti-de Sitter) groups. We can manage to have a mass term like $\overline{\Psi}\Psi$, as we have already seen. But, a term like $\overline{\Psi}\gamma^{\mu}\overrightarrow{D}_{\mu}\Psi$ in $(2+1)$ dimensions is more subtle. For the sake of simplicity, let us take a rigid transformation $g=e^{\frac{i}{2}\lambda^{AB}\mathbb{J}_{AB}}$ where $\lambda^{AB}$ are constants on the spacetime manifold. Then,
\begin{equation*}
\left(\overline{\psi}\gamma^{\mu}\partial_{\mu}\psi\right)'={E'}_{a}^{\mu}\overline{\Psi}'\gamma^{a}\partial_{\mu}\Psi'=\tensor{\Lambda}{_{a}^{c}}E_{c}^{\mu}\overline{\Psi}g^{-1}\gamma^{a}g\partial_{\mu}\Psi\;,
\end{equation*}
where we used \eqref{spinor_bar_transformation} and \eqref{spinor_transformation} in the second equality. As we observed above, the property  \eqref{gamma_as_vector_prop} ensures $g^{-1}\gamma^{a}g={\Lambda}{^{a}}{_{B}}\gamma^{B}$. But,
\begin{equation*}
\left(\overline{\Psi}\gamma^{\mu}\partial_{\mu}\Psi\right)'=\underbrace{\tensor{\Lambda}{_{a}^{c}}\tensor{\Lambda}{^{a}_{B}}}_{\mbox{is not a closed object}}E_{c}^{\mu}\overline{\Psi}\gamma^{B}\partial_{\mu}\Psi\;.
\end{equation*}
Note that if we were replaced $a$ with $A$ (requiring an extra dimension in all the theory as the vierbein $\tensor{e}{^{A}_{\mu}}$ must be well defined and the metric tensor accordingly), then $\tensor{\Lambda}{_{A}^{c}}\tensor{\Lambda}{^{A}}{_{B}}=\tensor{\delta}{^{c}_{B}}$ and, therefore, we will have
\begin{equation*}
\left(\overline{\Psi}\gamma^{\mu}\overrightarrow{D}_{\mu}\Psi\right)'=\overline{\Psi}\gamma^{\mu}\overrightarrow{D}_{\mu}\Psi\;.
\end{equation*}
The argument can be easily generalized to local transformations using the property \eqref{covariant_derivative_transf}.

This representation independent result implies that a term like $\overline{\Psi}\gamma^{\mu}\overrightarrow{D}_{\mu}\Psi$ in the $(2+1)$ dimensional action is not invariant under $SO(3,1)$ or $SO(2,2)$ groups (not even globally invariant!).

Let us now focus on the Hermiticity of the invariant terms. We can check that
\begin{equation*}
(i\overline{\Psi}\gamma^{\mu}\partial_{\mu}\Psi)^{\dagger}=-iE_{a}^{\mu}\partial_{\mu}\overline{\Psi}\gamma^{a}\Psi=-i\overline{\Psi}\gamma^{\mu}\partial_{\mu}\Psi \;,
\end{equation*}
where in this equality we used the Hermitian property \eqref{Hermiticity_gamma} of the Dirac matrices. Therefore a conventional massless Dirac flat action written explicitly Hermitian in $(2+1)$ dimensions is
\begin{equation}\label{flat_action_2}
S=\frac{i}{2} \int d^{3}x |e|  \left(\overline{\Psi}\gamma^{\mu}\partial_{\mu}\Psi-\partial_{\mu}\overline{\Psi}\gamma^{\mu}\Psi\right) \;,
\end{equation}
where $|e|=\mbox{det}\left[\tensor{e}{^{a}_{\mu}}\right]$.
In this flat case, we are safe. If we integrate by parts \eqref{flat_action_2} we obtain
\begin{equation}\label{flat_action_3}
S=i \int d^{3}x |e| \overline{\Psi}\gamma^{\mu}\partial_{\mu}\Psi\;,
\end{equation}
up to a boundary term. Therefore, both actions \eqref{flat_action_2} and \eqref{flat_action_3} give us the same field equations, neglecting boundary considerations.

This procedure is not equally simple when the spacetime manifold is curved or has torsion, because invariance of the action requires a spin connection to be added, as we said earlier. The covariant derivative is given in \eqref{covariant_derivative}, therefore we have to be careful with the Hermiticity of such a term,
\begin{equation*}
(i\overline{\Psi}\gamma^{\mu}\overrightarrow{D}_{\mu}\Psi)^{\dagger}=-iE_{a}^{\mu}\partial_{\mu}\overline{\Psi}\gamma^{a}\Psi - E_{a}^{\mu} \overline{\psi}\mathbb{J}_{BC}\omega^{BC}_{\mu}\gamma^{a}\Psi=-i\overline{\Psi}\overleftarrow{D}_{\mu}\gamma^{\mu}\Psi\;,
\end{equation*}
where we used the Hermiticity properties \eqref{Hermiticity_gamma} and \eqref{Hermiticity_J_extended}. This implies the Hermitian action in curved spacetime may be written as
\begin{equation}\label{curved_action}
S=\frac{i}{2} \int d^{3}x |e| \left( \overline{\Psi}\gamma^{\mu}\overrightarrow{D}_{\mu}\Psi - \overline{\Psi}\overleftarrow{D}_{\mu}\gamma^{\mu}\Psi \right)\;,
\end{equation}
but this time we have to keep both terms, as integration by parts gives us extra terms when we vary to obtain the field equations (see \cite{Shapiro}).

Now comes our assumption that the extended connection is of the form $\omega^{a3}_{\mu}=\sqrt{\lambda}e^{a}_{\mu}$, where $\sqrt{\lambda}$ is a real (imaginary) number for de Sitter (anti-de Sitter) case, implying that
\begin{eqnarray}\label{covariant_derivative_mod}
\overrightarrow{D}_{\mu}\Psi&=&\partial_{\mu}\Psi+\frac{i}{2}\omega^{ab}_{\mu}\mathbb{J}_{ab}\Psi+i\sqrt{\lambda}e^{a}_{\mu}\mathbb{P}_{a}\Psi \;,\\ \overline{\Psi}\overleftarrow{D}_{\mu}&=&\partial_{\mu}\overline{\Psi}-\frac{i}{2} \overline{\Psi}\mathbb{J}_{ab}\omega^{ab}_{\mu}-i \overline{\Psi}\mathbb{P}_{a}\sqrt{\lambda}e^{a}_{\mu}\;, \nonumber
\end{eqnarray}
where $\mathbb{P}_{a}\equiv\mathbb{J}_{a3}$ are the extra \emph{translation} generators\footnote{We called $\mathbb{P}_{a}$ translation generators although, we know these are not really translation generators (diffeomorphism generators) unless the torsion tensor is zero \cite{Witten1988}.}.

Observe the action \eqref{curved_action} can be split as
\begin{equation}\label{curved_action2}
S=\frac{i}{2} \int d^{3}x |e| \left[ \overline{\Psi}\gamma^{\mu}\overrightarrow{\widetilde{D}}_{\mu}\Psi - \overline{\Psi}\overleftarrow{\widetilde{D}}_{\mu}\gamma^{\mu}\Psi +  i\sqrt{\lambda}\overline{\Psi}\left\{\gamma^{a},\mathbb{J}_{a3}\right\}\Psi\right]\;,
\end{equation}
where $\overrightarrow{\widetilde{D}}_{\mu}$ and $\overleftarrow{\widetilde{D}}_{\mu}$ are the covariant derivatives \eqref{covariant_derivative_mod} containing only the Lorentz spin connection part $\omega^{ab}_{\mu}$. But, we can see immediately from the definition of generators \eqref{def_J} that
\begin{equation}\label{property_Ja3}
\left\{\gamma^{a},\mathbb{J}_{a3}\right\}=\frac{i}{2}\left\{\gamma^{a},\gamma_{a}\gamma_{3}\right\}=\frac{i}{2}\gamma^{a}\gamma_{a}\gamma^{3}+\frac{i}{2}\gamma_{a}\gamma_{3}\gamma^{a}=0\;.
\end{equation}
Therefore, the mixing term we want to have proportional to $\sqrt{\lambda}$ is zero. This result is also independent on the explicit representation of the Dirac matrices.

\subsection{Going one dimension up}
\label{sec:action}

From the above, we learn that to have a sensible Dirac action in $(2+1)$ dimensions, with $SO(3,1)$ or $SO(2,2)$ invariance, there are two issues, independent from the explicit representation of the gamma matrices: invariance, and Hermiticity.

As for invariance, terms like $\overline{\Psi}\gamma^{\mu}\overrightarrow{D}_{\mu}\Psi$ are not invariant under $SO(3,1)$ or $SO(2,2)$. As for Hermiticity, if we extend the Lorentz group where the connection related to $\mathbb{P}_{a}$ generators are proportional to the dreibeins $e^{a}_{\mu}$ (see \eqref{covariant_derivative_mod}), the mixing terms are not present in the action. We stress here that, even if the procedure to obtain the generators $\mathbb{J}_{AB}$ through \eqref{def_J} is direct, once we have four Dirac matrices, we do not have much room to step aside of this road. This is because there is a fine tuning manipulation to obtain objects transforming as scalars and vectors under the extended group (see \ref{appendix_J}).

Therefore, one concludes that, if we want to construct a $(2+1)$-dimensional Hermitian action, locally invariant under the (A)dS, it cannot be done with a minimal coupling prescription like \eqref{covariant_derivative_mod}.

On the other hand, if we are in $(3+1)$ dimensions, we can construct a Lorentz invariant term relaxing the prescription \eqref{covariant_derivative_mod}, i.e.,
\begin{eqnarray}\label{covariant_derivative_mod2}
\overrightarrow{D}_{\mu}\Psi&=&\partial_{\mu}\Psi+\frac{i}{2}\omega^{ab}_{\mu}\mathbb{J}_{ab}\Psi+i\omega^{a3}_{\mu}\mathbb{P}_{a}\Psi =\partial_{\mu}\Psi+\frac{i}{2}\omega^{A B}_{\mu}\mathbb{J}_{A B} \Psi \;, \\
\overline{\Psi}\overleftarrow{D}_{\mu} &=& \partial_{\mu}\overline{\Psi}-\frac{i}{2} \overline{\Psi}\mathbb{J}_{ab}\omega^{ab}_{\mu}-i \overline{\Psi}\mathbb{P}_{a}\omega^{a3}_{\mu}
= \partial_{\mu}\overline{\Psi}-\frac{i}{2} \overline{\Psi}\mathbb{J}_{AB}\omega^{AB}_{\mu} \;, \nonumber
\end{eqnarray}
meaning we are not restricting $\omega^{a3}_{\mu}$ to be proportional to the dreibein $e^{a}_{\mu}$ in this case, and, as before, $A=(a,3)$, etc.

This is as it must be. The gauge field associated to the $\mathbb{P}_{a}$ generators, being torsion present, and being translations traded for rotations in the larger space, can no longer be the vielbein.

%Eventually, the action emerging from this analysis is a standard $(3+1)$-dimensional $SO(3,1)$-invariant action for a Dirac field theory on background with curvature and torsion
%\begin{equation}\label{3+1_curved_action}
%S=\frac{i}{2} \int d^{4}x \left[ \overline{\Psi}\gamma^{\mu}\overrightarrow{D}_{\mu}\Psi - \overline{\Psi}\overleftarrow{D}_{\mu}\gamma^{\mu}\Psi \right]\;,
%\end{equation}
%with the covariant derivative of (\ref{covariant_derivative_mod2}), and with a third spatial dimension, of a purely abstract nature. The practical use of this action for computations in graphene, is something that remains to be seen.

%***Some comments on the third dimension***

%%%%%%%%%%%%%%%%%%%%%%%%%%%%%%%END Spin 1/2 fermion minimally coupled to AdS gauge field%%%%%%%%%%%%%%%%%%%%

%%%%%%%%%%%%%%%%%%%%INTERNAL SYMMETRY WAY%%%%%%%%%%%%%%%%%%%%%%%%%%%%%%%%%

\section{High-energy-theory correspondence II: Internal symmetry approach and supersymmetry}
\label{sec:USUSY}
%\keywords{Quantum Gravity Phenomenology, Torsion, Einstein-Cartan Theory, Graphene Correspondence}

To take into account the two Dirac points description, in this Section, we shall consider a more conventional \emph{bottom-up approach}. As we mentioned before, we can accommodate the $a$ and $b$ sublattice operators in very different convenient ways in the Hamiltonian dexcribing the $\pi$ electrons (see \ref{appendix_Dirac_points}). Here we take the Dirac points $3$ and $6$ of Fig. \ref{fig:HexagonBrilloin}, but with the matrices $M_{-}=\left(
                                                                        \begin{array}{cc}
                                                                          -1 & 0 \\
                                                                          0 & 1 \\
                                                                        \end{array}
                                                                      \right)=-N_{-}$.
The resultant Hamiltonian is
\begin{equation}\label{flat_Hamiltonian}
H=-i\hbar v_{F} \int d^{2}x \left(\psi^{\dagger}_{+}\vec{\sigma}\cdot\vec{\nabla}\psi_{+}+\psi^{\dagger}_{-}\vec{\sigma}\cdot\vec{\nabla}\psi_{-}\right)\;,
\end{equation}
We can interpret the $\pm$ subindex as different \emph{color} internal index\footnote{However, some references also call this index the \emph{flavor} quantum number \cite{Gusynin2007}. For our proposes, we call it color number.} \cite{GONZALEZ1993771}. Let us call $\psi_{\pm}=\psi_{1}$ and $\psi_{-}=\psi_{2}$, therefore we can write \eqref{flat_Hamiltonian} as
\begin{equation}\label{flat_Hamiltonian_internal}
H=-i\hbar v_{F} \sum\limits_{i=1}^{i=2} \int d^{2}x \psi^{\dagger}_{i}\vec{\sigma}\cdot\vec{\nabla}\psi_{i}\;.
\end{equation}

\subsection{Invariance of the flat action}

A crucial observation is that the Hamiltonian \eqref{flat_Hamiltonian_internal} is invariant under the global $SU(2)$ gauge transformation\footnote{According to \cite{Gusynin2007}, the internal group is even bigger, namely the $U(2)$ group.}
\begin{equation}\label{internal_transformation}
\psi_{i}\to\psi'_{i}={\left(e^{-i\lambda^{I}\sigma_{I}}\right)}{_{i}^{j}}\psi_{j}\;,
\end{equation}
where the capital Latin index $I=1,2,3$ is in the adjoint representation of $SU(2)$. On the other hand, the transformation \eqref{internal_transformation} means that the index $i=1,2$ belongs to the fundamental representation of $SU(2)$.

We can construct now the action associated to the Hamiltonian \eqref{flat_Hamiltonian_internal} taking
\begin{equation}\label{gamma_matrices}
\gamma^{0}=\sigma^{3}\;,\;\gamma^{1}=i\sigma^{2}\;,\;\gamma^{2}=-i\sigma^{1}\;.
\end{equation}
The convention \eqref{gamma_matrices} leads to the usual Clifford algebra \eqref{Clifford_algebra} in $(2+1)$ dimensions with the Hermiticity property \eqref{Hermiticity_gamma}.

The Lorentz conjugate is now\footnote{Our metric signature is $\eta_{ab}=\mbox{diag}(+,-,-)$.}
\begin{equation}\label{gamma_bar}
\overline{\psi}^{i}_{\alpha}=\psi^{\dagger j}_{\beta}{(\gamma^{0})}{^{\beta}_{\alpha}}\delta^{i}_{j}\;.
\end{equation}
Note the different kind of indexes in the definition \eqref{gamma_bar}, $\alpha$, $\beta$ are spinorial indexes while $i$, $j$ are color indexes.
Taking into account \eqref{flat_Hamiltonian_internal}, \eqref{gamma_matrices} and \eqref{gamma_bar}, we end up with the flat Hermitian action
\begin{equation}\label{flat_action_internal_Hermitian}
S=\frac{i}{2}\hbar v_{F} \int d^{3}x \left( \overline{\psi}^{i}\gamma^{\mu}\partial_{\mu}\psi_{i} - \partial_{\mu}\overline{\psi}^{i}\gamma^{\mu}\psi_{i} \right) \;,
\end{equation}
where the summation on the internal color index $i$ is understood.

As we can recognize \eqref{flat_action_internal_Hermitian} as the flat Dirac action in $(2+1)$ dimension, it is invariant under global $SO(2,1)$ transformations. Indeed, defining the Lorentz generators as usual \eqref{def_J}, we can check this action is invariant.

All in all, Hermitian action \eqref{flat_action_internal_Hermitian} is invariant under global transformations of the group $SO(2,1)\times SU(2)$.

We see from here that this route is very different from the previous sections. In this case the time reversal symmetry between the two Dirac points, as well as parity relations, are hidden. This hiding is what customarily done in the standard internal symmetry approach. For simplicity, we follow the same approach here, but we do keep in mind the difference with a purely internal doublet.

\subsection{The curved action}

How does the action \eqref{flat_action_internal_Hermitian} look like in a curved spacetime with a non-trivial bundle background?
The first guess would be to promote the global symmetry $SO(2,1)\times SU(2)$ of \eqref{flat_action_internal_Hermitian} to a local one through a covariant derivative. So,
\begin{equation}\label{curved_action_internal}
S=\frac{i}{2}\hbar v_{F} \int d^{3}x |e| \left[ \overline{\psi}^{i}\gamma^{\mu}{(\overrightarrow{D}_{\mu})}{_{i}^{j}}\psi_{j} - \overline{\psi}^{i}{(\overleftarrow{D}_{\mu})}{_{i}^{j}}\gamma^{\mu}\psi_{j} \right]  \;,
\end{equation}
where the covariant derivatives are written explicitly with all the indexes
\begin{eqnarray}\label{covariant_derivative_internal}
{(\overrightarrow{D}_{\mu})}{_{i}^{j}}\psi_{j}&=&\partial_{\mu}\psi_{i}+\frac{i}{2}\omega^{ab}_{\mu}\mathbb{J}_{ab}\psi_{i}+iA^{I}_{\mu}{(\sigma_{I})}{_{i}^{j}}\psi_{j} \;,\\ \overline{\psi}^{j}{(\overleftarrow{D}_{\mu})}{_{j}^{i}}&=&\partial_{\mu}\overline{\psi}^{i}-\frac{i}{2} \overline{\psi}^{i}\mathbb{J}_{ab}\omega^{ab}_{\mu}-i\overline{\psi}^{j}{(\sigma_{I})}{_{j}^{i}}A^{I}_{\mu}\;. \nonumber
\end{eqnarray}
Here $\omega^{ab}$ is the spin-connection and $A^{I}_{\mu}$ is a (real) non-Abelian gauge connection.
The covariant derivative \eqref{covariant_derivative_internal}, along with \eqref{Hermiticity_J_extended}, guaranty the action \eqref{curved_action_internal} to be local invariant under $SO(2,1)\times SU(2)$ and also Hermitian.

In the action \eqref{curved_action_internal}, we can isolate the torsion tensor $\tensor{T}{^{a}_{\mu\nu}}$ contribution. Indeed, splitting the Lorentz connection in the torsionless plus contorsion part $\tensor{\omega}{^{a}_{\mu}}=\tensor{\widetilde{\omega}}{^{a}_{\mu}}+\tensor{\kappa}{^{a}_{\mu}}$, we have
\begin{equation}\label{curved_action_internal_split}
S=\frac{i}{2}\hbar v_{F} \int d^{3}x |e| \left[ \overline{\psi}^{i}\gamma^{\mu}{(\overrightarrow{\widetilde{D}}_{\mu})}{_{i}^{j}}\psi_{j} - \overline{\psi}^{i}{(\overleftarrow{\widetilde{D}}_{\mu})}{_{i}^{j}}\gamma^{\mu}\psi_{j}  -\frac{1}{2}\tensor{\epsilon}{_{a}^{bc}}\tensor{T}{^{a}_{bc}}\overline{\psi}^{i}\psi_{i} \right]  \;,
\end{equation}
where $\overrightarrow{\widetilde{D}}_{\mu}$ and $\overleftarrow{\widetilde{D}}_{\mu}$ are the covariant derivatives \eqref{covariant_derivative_internal}, based only on the torsionless connection $\tensor{\widetilde{\omega}}{^{a}_{\mu}}$, and we defined $\tensor{T}{^{a}_{bc}}\equiv\tensor{E}{_{b}^{\mu}}\tensor{E}{_{c}^{\nu}}\tensor{T}{^{a}_{\mu\nu}}$.

\subsection{Conical singularity (pentagon)}

\begin{figure}[H]
\begin{center}
\includegraphics[width=0.6\textwidth,angle=0]{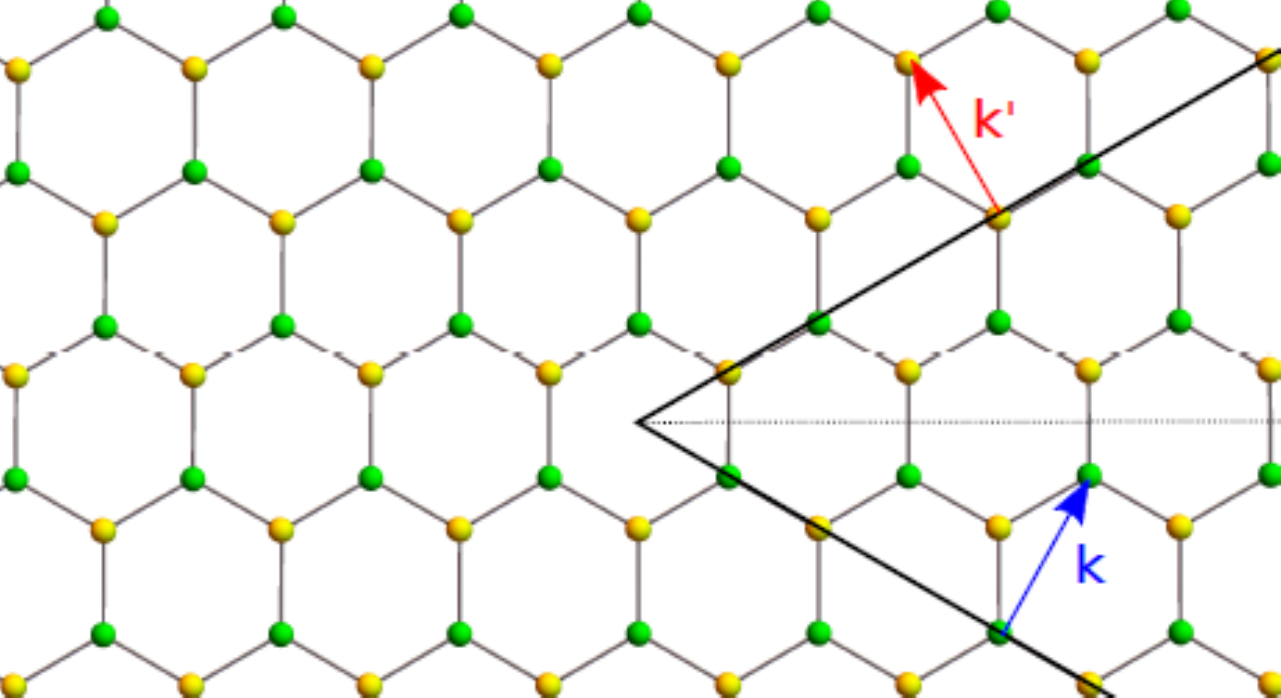}
\end{center}
\caption[3pt]{{\protect\small {{\bf Schematic view of the Volterra process for the creation of a pentagonal defect.}  From a flat lattice, the two black rays are identified, as well as the vectors $\vec{k}$ (blue) and $\vec{k'}$ (red). The resultant defect is depicted in Fig. \ref{fig:PentagonDoubleColoring}, along with the associated GGBs and consequential coloring patterns.}}}%
\label{pentagon}%
\end{figure}

As a Volterra process, the conical singularity can be realized by identifying two adjacent rays as in Fig. \ref{pentagon}.
We have a non-trivial spin-connection $\omega^{ab}_{\mu}$ related to the rotation group when the two lines are identified. In this case, $k_{x}\to-k_{x}$ and $k_{y}\to k_{y}$, which due to the lattice symmetries is equivalent to $\vec{k}\to-\vec{k}$ (see Fig. \ref{pentagon}). This effect can be generated with $\mathbb{J}_{12}=\sigma^{3}$ and a spin connection proportional to $1/r$.

It can be shown that under a conical singularity, the two Dirac points are interchanged \cite{GONZALEZ1993771}. The way to mimic this effect is through a \emph{fictitious} magnetic flux emerging from the apex of the cone. In this way, the non-Abelian internal gauge field can be written in this case as $A=A^{(2)}_{\theta}\sigma^{2}$, where $A^{(2)}_{\theta}=\frac{g}{2\pi r}$ and $g$ is a magnetic charge to be tuned by the experiments. This is the building block to construct the field equations to describe curved graphene sheets (tetrahedron, dodecahedron or $C_{60}$ fullerenes), where different magnetic flux are added for each vertex which contain a color line frustration, pointing out to a magnetic monopole at the center of the molecule structure \cite{GONZALEZ1993771}.

For a heptagon, the situation is very similar, but with opposite sign for the charge $g$.

\subsection{Pair of defects (heptagon–pentagon) and many defects (grain boundaries) generalization}

\begin{figure}[H]
\begin{center}
\includegraphics[width=0.5\textwidth,angle=0]{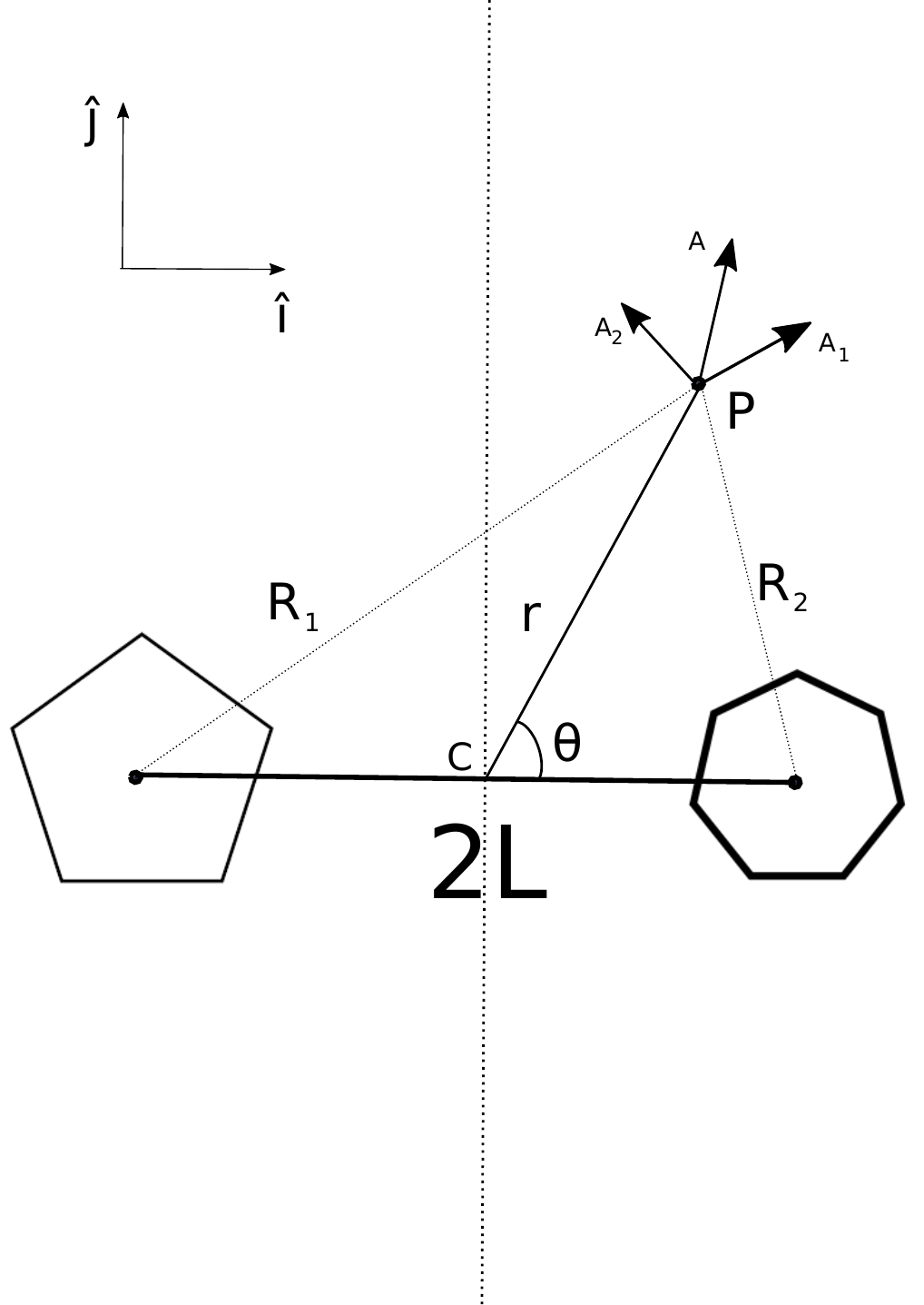}
\end{center}
\caption[3pt]{{\protect\small {{\bf Internal gauge field created by a heptagon–pentagon pair}.  The fields generated by the pentagon $\vec{A}_{1}$ and heptagon $\vec{A}_{2}$ are perpendicular to the segment joining the defect with the point $P$, respectively. Note that the magnetic flux $g$ created by the pentagon is opposite to the heptagon, $-g$.  The total field on $P$ is $\vec{A}^{(2)}=\vec{A}_{1} + \vec{A}_{2}$, which depends on the distance $r$ from $C$ to $P$, the angle $\theta$ and also on the distance $2L$ between defects (see equation \eqref{gauge_two_singularities}).}}}%
\label{two_singularities}%
\end{figure}

\begin{figure}[H]
\begin{center}
\includegraphics[width=0.5\textwidth,angle=0]{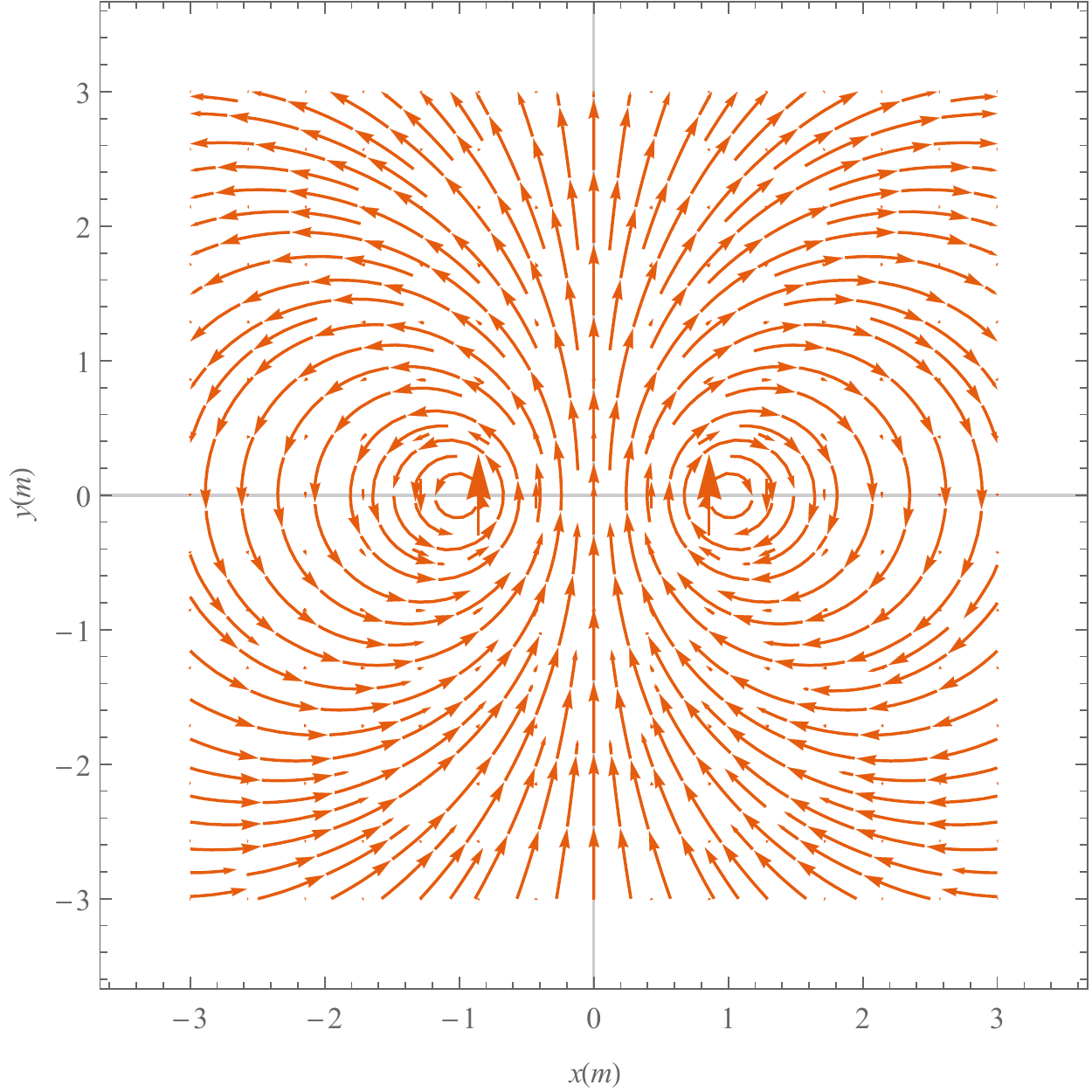}
\end{center}
\caption[3pt]{{\protect\small {{\bf Gauge field created by a heptagon–pentagon pair.} The pentagon is localized at $(-L,0)$, while the heptagon at $(0,L)$. For the plot, we take $L=1\mbox{ m}$ and $g=1$. The lines represent the direction of the internal gauge field $\vec{A}^{(2)}$ given by the formula \eqref{gauge_two_singularities} in different points of a hypothetical $6\mbox{ m}$ square sheet of graphene (or graphene-like) material. The conventions for angles and distances are the same as Fig. \ref{two_singularities}.}}}%
\label{heptagon-pentagon_gauge}%
\end{figure}

We can realize the situation with heptagon–pentagon defects (similar to Fig. \ref{fig:edge-dislocation}) separated by a distance $2L$ as the problem to find a magnetic field in a plane perpendicular to opposite wire currents. A sketch of the problem is given in Fig. \ref{two_singularities}. The total gauge field $\vec{A}=\vec{A}^{(2)}\sigma^{2}$ on the point $P$ depends on the distance $r$ to the midpoint of the segment joining the pentagon and heptagon centers $C$, and the angle $\theta$ of with respect to such a line (see Fig. \ref{two_singularities} for details). The result is
\begin{equation}\label{gauge_two_singularities}
\vec{A}^{(2)}=\frac{g}{2\pi}\left[r\sin\theta\left(-\frac{1}{R_{1}^{2}}+\frac{1}{R_{2}^{2}}\right)\widehat{i}+\left(\frac{L+r\cos\theta}{R_{1}^{2}}+\frac{L-r\cos\theta}{R_{2}^{2}}\right)\widehat{j}\right]\;,
\end{equation}
where $R_{1}^{2}=L^{2}+r^{2}+2Lr\cos\theta$ and $R_{2}^{2}=L^{2}+r^{2}-2Lr\cos\theta$. A sketch of the gauge field for this case is given in Fig. \ref{heptagon-pentagon_gauge}.

When $r\gg L$, then
\begin{equation*}
\vec{A}^{(2)}=\frac{-gL}{\pi r^{2}}\left[\sin2\theta\widehat{i}+\cos2\theta\widehat{j}\right]\;.
\end{equation*}
This makes sense, when the defect is far enough, the gauge field decreases as $1/r^{2}$ meaning that loop integral around the center of the pair of defects $C$, decreases as $1/r$. The latter explains why the heptagon-pentagon defect does not mix the Dirac points when these defects are very localized.

To take into account the many defects case, we generalize the previous case, by adding more and more pairs of heptagon-pentagon defects.% (see figure \ref{grain_boundaries}).

%\begin{figure}
%\begin{center}
%\includegraphics[width=0.5\textwidth,angle=0]{grain_boundaries.pdf}
%\end{center}
%\caption[3pt]{{\protect\small {Generalization description of two defect could emerge the GB situation.}}}%
%\label{grain_boundaries}%
%\end{figure}

\subsection{The unconventional supersymmetry as a more general theory}

Remarkably, the action \eqref{curved_action_internal_split} is very similar to the action of USUSY for an external non-abelian $SU(2)$ gauge field and a fixed curved background \cite{u-susy-su2}. In such work, the $SU(2)$ label represented the spin degree of freedom, but here this $SU(2)$ internal gauge field has a direct interpretation as the field which mix the Dirac points $K_{+}$ and $K_{-}$.

In this USUSY model, we take the one-form connection spanned by the Lorentz generators $\mathbb{J}_{a}$, the $SU(2)$ generators corresponding to the internal gauge symmetry $\mathbb{T}_{I}$, the supercharges $\overline{\mathbb{Q}}^{i}$ and $\mathbb{Q}_{i}$ (note that these last generators contains the index corresponding to the fundamental group of $SU(2)$ as well as the spinors)\footnote{It is possible to add a central extension generator $\mathbb{Z}$ and its corresponding one-form coefficient $b$ \cite{u-susy-su2}. However, we shall not consider this extension in the present work.} \cite{u-susy-su2}
\begin{equation}\label{conncetion_su(2)}
\mathbb{A}=A^{I}\mathbb{T}_{I}+\overline{\psi}^{i}\slashed{e}\mathbb{Q}_{i}+\overline{\mathbb{Q}}^{i}\slashed{e}\psi_{i}+\omega^{a}\mathbb{J}_{a}\;,
\end{equation}
where $A^{I}=A^{I}_{\mu}dx^{\mu}$ is the one-form $SU(2)$ connection, $\omega^{a}=\tensor{\omega}{^{a}_{\mu}}dx^{\mu}$ is the one-form Lorentz connection in $(2+1)$ dimensions, and we defined the one-form $\slashed{e}\equiv\tensor{e}{^{a}_{\mu}}\gamma_{a}dx^{\mu}$.
We can construct a three-form Chern-Simons action from \eqref{conncetion_su(2)}, namely\footnote{Here, we omitted the wedge notation for the exterior product. For instance, $\mathbb{A}^{3}$ stands for the three-form $\mathbb{A}\wedge\mathbb{A}\wedge\mathbb{A}$.}
\begin{equation}\label{action_su(2)}
L=\frac{\kappa}{2}\langle\mathbb{A}d\mathbb{A}+\frac{2}{3}\mathbb{A}^{3}\rangle\;,
\end{equation}
where $\langle\ldots\rangle$ is the invariant supertrace of $\mathfrak{usp}(2,1|2)$ graded Lie algebra and $\kappa$ is a dimensionless constant. This way, the Lagrangian can be written simply as
\begin{equation}\label{action_su(2)_expanded}
L=\frac{\kappa}{4}\left(A^{I}dA_{I}+\frac{1}{3}\epsilon_{IJK}A^{I}A^{J}A^{K}\right)+\frac{\kappa}{4}\left(\omega^{a}d\omega_{a}+\frac{1}{3}\epsilon_{abc}\omega^{a}\omega^{b}\omega^{c}\right)+L_{\psi}\;,
\end{equation}
where the fermionic part is
\begin{equation*}
L_{\psi}=\kappa\overline{\psi}\left(\gamma^{\mu}\overrightarrow{D}_{\mu}-\overleftarrow{D}_{\mu}\gamma^{\mu}-\frac{1}{2}\tensor{\epsilon}{_{a}^{bc}}\tensor{T}{^{a}_{bc}}\right)\psi|e|d^{3}x\;.
\end{equation*}
We can see the action \eqref{action_su(2)_expanded} possesses also a local scale (Weyl) symmetry. Indeed, by scaling the vierbein and the fermions as
\begin{equation*}
\tensor{e}{^{a}_{\mu}} \to \tensor{e}{^{a}_{\mu}}'=\lambda\tensor{e}{^{a}_{\mu}} \; , \; \psi \to \psi'=\lambda^{-1}\psi \;,
\end{equation*}
where $\lambda=\lambda(x)$ is a non-singular function on the spacetime manifold, the action \eqref{action_su(2)_expanded} is invariant. This is a consequence of the particular construction of the connection \eqref{conncetion_su(2)}, where the fermion always appear along with the vierbein field, forming a composite field.

If the geometric background is fixed and the non-Abelian gauge field is external (there is no dynamics for the phonons and gauge fields), then the Lagrangian \eqref{action_su(2)_expanded} leads to the following action
\begin{equation}\label{action_su(2)_fermions}
S_{\psi}=\kappa\int\overline{\psi}\left(\gamma^{\mu}\overrightarrow{\widetilde{D}}_{\mu}-\overleftarrow{\widetilde{D}}_{\mu}\gamma^{\mu}-\frac{1}{4}\tensor{\epsilon}{_{a}^{bc}}\tensor{T}{^{a}_{bc}}\right)\psi|e|d^{3}x\;.
\end{equation}
Therefore, we can see that the only difference of $S_{\psi}$ with respect to \eqref{curved_action_internal_split} is the coefficient in front of the torsion term. The role of torsion in USUSY is to give an effective mass to the electrons (without mixing the Dirac points), very similar to what happens in the action \eqref{curved_action_internal_split}.

A peculiarity of USUSY is the absence of gravitini, although it includes gravity and supersymmetry. Likewise, no gauginos are present. All the parameters involved in the system are either protected by gauge invariance or emerge as integration constants. Interesting enough, the vacuum sector is defined by configurations with locally flat Lorentz and $SU(2)$ connections carrying nontrivial global charges, as is the case of BTZ black holes \cite{BTZ1992}. Moreover, the only propagating degrees of freedom are the fermionic ones \cite{GPZ}.

%%%%%%%%%%%%%%%%%%%%%%%%END INTERNAL SYMMETRY WAY%%%%%%%%%%%%%%%%%%%%%%

\section{Conclusions and Discussion}
\label{sec:conc}

We have probed two alternative quantum field theoretical emergent scenarios corresponding to grain boundaries, especially in graphene (but also in related materials), in a continuum, low-energy description. Crucial is the necessity of having two Dirac points, hence two copies of the Dirac Hamiltonian. As we have learned already \cite{IorioPais}, also here the structure of the $k$-space in graphene plays a crucial role in determining the topology/geometry of the emergent fields, both background classical ``spacetimes'', and ``matter fields'' on those backgrounds.

In the spatiotemporal scenario, we explored the necessity for an extension of the geometric/relativistic corresponding gauge group. We found that this is possible, but the most natural setting is to move from the Lorentz group in three dimensions $SO(2,1)$ to one spatial dimension up, namely $SO(3,1)$. On the one hand, this result is fascinating, as it points to a Dirac theory in $(3+1)$ dimensions, with a symmetry group, $SO(3,1)$, locally isomorphic to the Lorentz group there. On the other hand, though, the operational meaning of such $(3+1)$-dimensional spacetime, with an abstract $\tilde{z}$ third coordinate, the role and meaning of the ``cosmological constant'', and how this could be used to extract physically meaningful prediction, are still to be clarified.

It also needs clarification the role of torsion involved here. For instance, since we have here three space dimensions, it could be possible to overcome, in this corresponding spacetime, the obstructions encountered in the literature \cite{Mesaros:2009az,Cortijo:2006xs,deJuan2010,Vozmediano:2010zz,Amorim:2015bga}. On the other hand, we plan to explore in a forthcoming work a different and more conservative approach, keeping the number of dimensions fixed to $(2+1)$, and introducing torsion through the standard Burger vector, but including time components of the torsion tensor.

Let us recall that general relativity has no torsion, and the latter is not seen in experiments. Nonetheless, it is intimately connected with the existence of spinors, in various contexts, see, e.g., \cite{KibbleTorsion1961, Shapiro, Hehl:1976kj, CartanTheoryHehl, PercacciTorsion}, including standard supersymmetry \cite{WessBagger}. This was a motivation for the first investigation, but also motivated our second approach, the internal symmetry one. There we focused on the less unusual scenario that links the doublet of Dirac points to an internal symmetry, but we offered here an analysis that links it with USUSY \cite{AVZ}.  Supersymmetry neither is seen in experiments, hence our findings here should be taken as the possibility to reproduce such scenarios in an indirect way.

With this in mind, it is then important to look for corresponding (analogue) systems where both phenomena, torsion and supersymmetry, might be reproduced, so to learn from there about the fundamental open issues.

\section*{Acknowledgments}

The authors thank Fabrizio Canfora, Francisco Correa, Gast\'on Giribet, Guillermo Silva, Adamantia Zampeli, and Jorge Zanelli for inspirational discussions. A.~I. also gladly acknowledges the warm hospitality of the Centro de Estudios Cientificos (CECs) of Valdivia, Chile, and of the Departments of Physics of the Universities of Trento and Salerno, Italy, while parts of this paper were conceived. P.~P. was supported by a PDM grant from KU Leuven.

\appendix

\section{Torsion and Grain Boundaries}
\label{appendix_torsionGB}

There are several issues related to torsion in two-dimensional layer systems. We mention here the most relevant for our purposes. First, even though torsion is related to the Burger vector as in \eqref{torsion-Burgers}, there are technical problems to express analytically torsion in some edge dislocation region once the Burger vector is given on it \cite{Lazar2003}. Second, according to the Frank formula, we cannot associate a Burger vector $\vec{b}$ \emph{only} with a misorientation angle $\theta$, but we need another piece of information from the lattice, namely the distance $d$ between dipole defects \cite{Yazyev2010}. To be concrete, the Frank formula for small $\theta$ angles is (see \cite{hirth1967theory,Yazyev2010})
\begin{equation}\label{Frank_formula}
\theta=2\arcsin\left(\frac{|\vec{b}|}{2d}\right)\;.
\end{equation}
For a given $\theta$, we can associate an infinite number of Burger vectors, as we can tune the distance between dipoles dislocations $d$ to obtain a misorientation angle $\theta$. So, according to \eqref{torsion-Burgers}, do we have an infinite number of torsion tensors for a given $\theta$?

In fact, torsion is not directly related to true GBs. Of course, in the lattice, indirectly, all the defects are related, e.g., a pair of disclinations results in a dislocation, as is shown in the Fig. \ref{fig:edge-dislocation}. Nonetheless, the continuum limit of these defects may be a different matter. The point is that the quantity associated to GBs is $\theta$, which has a different geometrical nature than $\vec{b}$ or than disclination angle $s$. This is sketched in Table \ref{table_defects}.
\begin{table}
      \centering
      \begin{tabular}{|l|l|l|l|}
      \hline
      Level & Defect & Lattice description & Continuous description \\
      \hline
      $1$ & Disclination & $s$ & Curvature \\
      $2$ & Dislocation  & $\vec{b}$ & Torsion \\
      $3$ & Grain Boundary & $\theta$ & Parity Transformation \\
      \hline
    \end{tabular}\caption{Characterization of the defects in a two-dimensional lattice and their continuum limit. Each level of defect in the hierarchy has different quantities that characterize them.}\label{table_defects}
    \end{table}
In this approach, there is a kind of hierarchy \cite{Yazyev2010,Yazyev2014}:
\begin{itemize}
\item The first level is a single disclination characterized by the disclination angle $s$ in the lattice and by curvature in the continuum limit
\item Second level, from two disclinations of opposite $s$ angles, we construct a dislocation characterized by the Burger vector $\vec{b}$ in the lattice (as shown in Fig. \ref{fig:edge-dislocation}), and by torsion in the continuum limit. But, we cannot define an angle $s$ for this defect, therefore there is no a defined curvature
\item Third level, from a distribution of dislocations, each one with $\vec{b}$ separated by a distance $d$, we can construct a GB characterized by $\theta$. But, we cannot define a single $\vec{b}$ for such a defect, therefore there is not a defined torsion
\end{itemize}
From one side, we can see that disclinations, dislocations and GBs are all related, but at same time in each step certain combinations of them create different kind of defects characterized by different topological invariants: $s$, $\vec{b}$ and $\theta$, respectively (see Table \ref{table_defects}).

\section{Two Dirac points representations}
\label{appendix_Dirac_points}

In this paper we stop at the near neighbor approximation, for which the tight-binding Hamiltonian is
\begin{equation}\label{Hfirst}
    H = \sum_{\vec{k}} \left( {\cal F} (\vec{k}) a^\dag_{\vec{k}} b_{\vec{k}} + {\cal F}^* (\vec{k}) b^\dag_{\vec{k}} a_{\vec{k}} \right)  \;,
\end{equation}
with ${\cal F} (\vec{k}) = - \eta e^{- i k_y} [1 + 2 e^{i \frac{3}{2} k_y} \cos(\frac{\sqrt{3}}{2} k_x)]$. Solutions of ${\cal F} = 0$ are the Dirac points, indicated in Fig. \ref{fig:HexagonBrilloin} for the armchair choice of the Brillouin zone. With reference to the figure, choosing point $1$ there as the first Dirac point, the other, nonequivalent Dirac point can be taken to be 4 (corresponding to a full inversion $\vec{k} \to - \vec{k}$), or $2$ (corresponding to a parity transformation $k_x \to - k_x$), or $6$ (corresponding to a rotation of $\pi / 3$). Other choices of the Brillouin zone, see, e.g., \cite{PacoReview2009}, give a zigzag cell, related to the former by a rotation matrix
$
\left(                                                                                                                                                                                                                       \begin{array}{cc}
\sqrt{3}/2 & - 1 \\
1 & \sqrt{3}/2 \\
\end{array}
\right)
$.
This way the pair $1$-$4$ above turns into  $1'$-$4'$ (corresponding to a parity transformation, but on the $y$ axis, $k_y \to - k_y$), 1-2 turns into 1'-2' (corresponding to a rotation of $\pi / 3$), 1-6 turns into $1'$-$6'$
(corresponding to $k_y \to - k_y$).

Let us now give in Table \ref{table_representations} the possible choices of the Dirac-like Hamiltonian (second column), that amounts to the possible choices of the two inequivalent Dirac points (first column). The third column reports what type of transformation relates the two Dirac points, while the last column indicates in which cases the latter transformation can be suitably described by the same transformation in the Dirac language (e.g., applying an $x$-parity transformation to the $H$ of the $1$-$2$ case one indeed obtains invariance of the combined components, say, $H_{+}$ and $H_{-}$). As said in the main text, this means that in certain variables one can do certain considerations more explicitly, and associate a proper parity change to the change of sublattice (coloring-pattern), the latter can be associated to a interchange of Dirac points, and, since the actual condensed matter physics is independent from this choice, we can always pick up a description where this is true.
\begin{table}
\begin{tabular}{|c|c|c|c|}
  \hline
  Dirac points & $H$ structure & Transformation & Suitability \\ \cline{1-4}
  1 - 2  & ${\tilde{\psi}}_+^\dag \vec{\sigma^*} \cdot \vec{p} {\tilde{\psi}}_+ - {\tilde{\psi'}}_-^\dag \vec{\sigma} \cdot \vec{p} {\tilde{\psi}'}_-$& $x$-parity & yes \\
  1 - 4  & ${\tilde{\psi}}_+^\dag \vec{\sigma^*} \cdot \vec{p} {\tilde{\psi}}_+ - {\tilde{\tilde{\psi}}'}_-^\dag \vec{\sigma} \cdot \vec{p} {\tilde{\tilde{\psi}}'}_-$ & full inversion  &  \\
  1 - 6  & ${\tilde{\psi'}}_-^\dag \vec{\sigma} \cdot \vec{p} {\tilde{\psi}'}_- + \psi_+^\dag \vec{\sigma} \cdot \vec{p} \psi_+$ & $\pi/3$  rotation &  \\
  2 - 3  & $- {\tilde{\psi}}_+^\dag \vec{\sigma^*} \cdot \vec{p} {\tilde{\psi}}_+ - {\psi'}_-^\dag \vec{\sigma} \cdot \vec{p} {\psi'}_-$ & $\pi/3$  rotation & \\
  2 - 5  & $- {\tilde{\psi}_+}^\dag \vec{\sigma^*} \cdot \vec{p} {\tilde{\psi}_+} + {\tilde{\tilde{\psi}}'}_-^\dag \vec{\sigma} \cdot \vec{p} {\tilde{\tilde{\psi}}'}_-$ & full inversion & \\
  3 - 4  & $- \psi_+^\dag \vec{\sigma^*} \cdot \vec{p} \psi_+  - {\tilde{\tilde{\psi}}'}_-^\dag \vec{\sigma} \cdot \vec{p} {\tilde{\tilde{\psi}}'}_-$ & $\pi/3$  rotation &  \\
  3 - 6  & $- {\psi'}_-^\dag \vec{\sigma} \cdot \vec{p} {\psi'}_- + \psi_+^\dag \vec{\sigma} \cdot \vec{p} \psi_+ $ & full inversion & yes \\
  4 - 5  & $- {\tilde{\tilde{\psi}}}_+^\dag \vec{\sigma^*} \cdot \vec{p} {\tilde{\tilde{\psi}}}_+ + {\tilde{\tilde{\psi}}'}_-^\dag \vec{\sigma} \cdot \vec{p} {\tilde{\tilde{\psi}}'}_-$ & $x$-parity & yes \\
  5 - 6  & ${\tilde{\tilde{\psi}}'}_-^\dag \vec{\sigma} \cdot \vec{p} {\tilde{\tilde{\psi}}'}_- + \psi_+^\dag \vec{\sigma} \cdot \vec{p} \psi_+$ & $\pi/3$ rotation & \\
  \hline
\end{tabular}
\caption[Text excluding the matrix]{Possible choices of the two inequivalent Dirac points to write down the Hamiltonian describing the $\pi$ electrons. Here $\psi  = \left(\begin{array}{c}
          b \\
          a
        \end{array} \right)$,
$\tilde{\psi}  =  \left( \begin{array}{c}
          e^{i \pi/3} b \\
          a
        \end{array} \right)$,
$\tilde{\tilde{\psi}} = \left( \begin{array}{c}
          e^{- i \pi/3} b \\
          a
        \end{array} \right)$,
$\psi'  = \left( \begin{array}{c}
          a \\
          b
        \end{array} \right)$ ,
$\tilde{\psi'} =  \left( \begin{array}{c}
          e^{i \pi/3} a \\
          b
        \end{array} \right)$,
$\tilde{\tilde{\psi''}}  = \left( \begin{array}{c}
          e^{- i \pi/3} a \\
          b
        \end{array} \right)$, and, as usual $\vec{\sigma} = (\sigma_1, \sigma_2)$ and $\vec{\sigma^*} = (\sigma_1, - \sigma_2)$.}\label{table_representations}
\end{table}

It is possible to obtain many other representations from each row of the Table \ref{table_representations}. Indeed, taking two $2\times2$ matrices $M_{\pm}$ and $N_{\pm}$, such that $M_{\pm}N_{\pm}=\pm\mathbb{I}_{2\times2}$, then re-defining $\psi'=N_{\pm}\psi$ and $\vec{\sigma'}=M_{\pm}^{\dagger}\vec{\sigma}M_{\pm}$, we have $\psi_{\pm}'^{\dagger}\sigma'\psi_{\pm}'=\psi_{\pm}^{\dagger}\sigma\psi_{\pm}$.

For instance, the Hamiltonian $H$ associated to the Dirac points $3$ and $6$ in Table \ref{table_representations} can be re-written if we take  $M_{-}=\left(
                                                                        \begin{array}{cc}
                                                                          0 & 1 \\
                                                                          1 & 0 \\
                                                                        \end{array}
                                                                      \right)=N_{-} \;.$
In this way, we obtain a couple of massless Dirac fermions in a two-dimensional flat manifold \cite{IorioReview}
\begin{equation*}
H=-i\hbar v_{F} \int d^{2}x \left(\psi^{\dagger}_{+}\vec{\sigma}\cdot\vec{\nabla}\psi_{+}-\psi^{\dagger}_{-}\vec{\sigma}^{*}\cdot\vec{\nabla}\psi_{-}\right)\;,
\end{equation*}
with the bi-spinors defined as $\psi_{\pm}=\left(
                                  \begin{array}{c}
                                    b_{\pm} \\
                                    a_{\pm} \\
                                  \end{array}
                                \right)
$.

Another option, which is used in Section \ref{sec:USUSY}, is to take for the same $3$ and $6$ Dirac points, but with $M_{-}=\left(
                                                                        \begin{array}{cc}
                                                                          -1 & 0 \\
                                                                          0 & 1 \\
                                                                        \end{array}
                                                                      \right)=-N_{-}$,
where the resultant Hamiltonian is
\begin{equation*}
H=-i\hbar v_{F} \int d^{2}x \left(\psi^{\dagger}_{+}\vec{\sigma}\cdot\vec{\nabla}\psi_{+}+\psi^{\dagger}_{-}\vec{\sigma}\cdot\vec{\nabla}\psi_{-}\right)\;,
\end{equation*}
and, in this case,
\begin{equation*}
\psi_{+}=\left(
           \begin{array}{c}
             b_{+} \\
             a_{+} \\
           \end{array}
         \right)               \; \;     \psi_{-}=\left(
           \begin{array}{c}
             a_{-} \\
             -b_{-} \\
           \end{array}
         \right)        \;.
\end{equation*}

\section{Generators $\mathbb{J}_{AB}$}
\label{appendix_J}

We can construct the $\mathbb{J}_{AB}$ generators with the algebra (\ref{AdS_algebra}), in many different ways \cite{Z-book}. However, here we show that if we want they fulfill property \eqref{gamma_as_vector_prop} (implying that $\overline{\psi}\gamma^{A}\psi$ be a vector), these generators must be of the form \eqref{def_J}.

\begin{lemma}
  If $\mathbb{J}_{AB}$ are six generators antisymmetric in the $A,B=0,1,2,3$ indexes, then
  $J_{AB}=\frac{i}{4}[\gamma_{A},\gamma_{B}]$ $\Leftrightarrow$ $\left[\gamma^{C},\mathbb{J}_{AB}\right]=i\left(\delta^{C}_{A}\gamma_{B}-\delta^{C}_{B}\gamma_{A}\right)$.
\end{lemma}

Proof:

  $(\Rightarrow)$

  If $J_{AB}=\frac{i}{4}[\gamma_{A},\gamma_{B}]$ then
  \begin{eqnarray*}
    \left[\gamma^{C},\mathbb{J}_{AB}\right]&=&\left[\gamma^{C},\frac{i}{4}[\gamma_{A},\gamma_{B}]\right]=\frac{i}{4}\left(\gamma^{C}[\gamma_{A},\gamma_{B}]-[\gamma_{A},\gamma_{B}]\gamma^{C}\right)\\
    &=&\frac{i}{4}\left\{\gamma^{C},\gamma_{A}\right\}\gamma_{B}-\frac{i}{4}\gamma_{A}\left\{\gamma^{C},\gamma_{B}\right\}-\frac{i}{4}\left\{\gamma^{C},\gamma_{B}\right\}\gamma_{A}+\frac{i}{4}\gamma_{B}\left\{\gamma^{C},\gamma_{A}\right\}\\
    &=&i\left(\delta^{C}_{A}\gamma_{B}-\delta^{C}_{B}\gamma_{A}\right) \;,
  \end{eqnarray*}
  where we used in the second line the property $[C,AB]=\{C,A\}B-A\{C,B\}$ for $A,B,C$ arbitrary matrices, and also the Clifford algebra \eqref{Clifford_algebra} in the third line.

  $(\Leftarrow)$

  We have already shown that $[\gamma^{C},\frac{i}{4}[\gamma_{A},\gamma_{B}]]=i\left(\delta^{C}_{A}\gamma_{B}-\delta^{C}_{B}\gamma_{A}\right)$, therefore $[\gamma^{C},\frac{i}{4}[\gamma_{A},\gamma_{B}]-\mathbb{J}_{AB}]=0$. Let us define the matrices $D_{AB}=\frac{i}{4}[\gamma_{A},\gamma_{B}]-\mathbb{J}_{AB}$, then $\left[\gamma^{C},D_{AB}\right]=0$. As by hypothesis $\mathbb{J}_{AB}$ is antisymmetric in $A,B$ indexes, then also $D_{AB}=-D_{BA}$. The only way to match the indexes in four dimensions (apart from a term proportional to $\frac{i}{4}[\gamma_{A},\gamma_{B}]$) is $D_{AB}\propto{\epsilon}{_{AB}}{^{FG}}\gamma_{F}\gamma_{G}=\frac{1}{2}{\epsilon}{_{AB}}{^{FG}}[\gamma_{F},\gamma_{G}]$. But,
   \begin{eqnarray*}
    \frac{1}{2}{\epsilon}{_{AB}}{^{FG}}\left[\gamma^{C},[\gamma_{F},\gamma_{G}]\right]&=&-2i{\epsilon}{_{AB}}{^{FG}}\left[\gamma^{C},\frac{i}{4}[\gamma_{F},\gamma_{G}]\right]\\
    &=&2{\epsilon}{_{AB}}{^{CG}}\gamma_{G}-2{\epsilon}{_{AB}}{^{FC}}\gamma_{F}\\
    &=&4{\epsilon}{_{AB}}{^{CF}}\gamma_{F} \;,
    \end{eqnarray*}
  where we used again the hypothesis in the second equality.
  This term is in general nonzero, contradicting the fact that by construction $\left[\gamma^{C},D_{AB}\right]=0$. Therefore, the only possible is $D_{AB}=0$, implying that $\mathbb{J}_{AB}=\frac{i}{4}[\gamma_{A},\gamma_{B}]$.

\rightline {\it Q.E.D.}

One explicit representation for these generators is the following. Take the Dirac matrices as
\begin{eqnarray}\label{gamma_explicit}
\gamma^{0}=\left(
             \begin{array}{cc}
               \sigma^{3} & 0 \\
               0 & \sigma^{3} \\
             \end{array}
           \right) \;&,&
\gamma^{1}=\left(
             \begin{array}{cc}
               i\sigma^{2} & 0 \\
               0 & -i\sigma^{2} \\
             \end{array}
           \right) \; , \nonumber \\ \; \gamma^{2}=\left(
             \begin{array}{cc}
               -i\sigma^{1} & 0 \\
               0 & -i\sigma^{1} \\
             \end{array}
           \right) \; &,&
           \gamma^{3}=\left(
             \begin{array}{cc}
               0 & \sigma^{2} \\
               -\sigma^{2} & 0 \\
             \end{array}
           \right) \; ,
\end{eqnarray}
which satisfy the Clifford algebra \eqref{Clifford_algebra} for $\eta^{AB}=\mbox{diag}(+,-,-,-)$.
The direct way to obtain the $\mathbb{J}_{AB}$ is by using the formula \eqref{def_J},
{\small
\begin{eqnarray}\label{J_and_P_matrices}
\mathbb{J}_{01}&=&\frac{i}{2}\left(
                    \begin{array}{cc}
                    -\sigma^{1} & 0 \\
                    0 & \sigma^{1} \\
                    \end{array}
                    \right)                   \; ,
\mathbb{J}_{02}=-\frac{i}{2}\left(
                    \begin{array}{cc}
                    \sigma^{2} & 0 \\
                    0 & \sigma^{2} \\
                    \end{array}
                    \right)                   \; ,
\mathbb{J}_{12}=\frac{1}{2}\left(
                    \begin{array}{cc}
                     \sigma^{3} & 0 \\
                     0 & -\sigma^{3} \\
                    \end{array}
                    \right)                   \; , \\
\mathbb{J}_{03} &=&\frac{1}{2}\left(
                    \begin{array}{cc}
                    0 & -\sigma^{1} \\
                    \sigma^{1} & 0 \\
                    \end{array}
                    \right)                   \; ,
\mathbb{J}_{13} =-\frac{1}{2}\left(
                    \begin{array}{cc}
                    0 & \mathbb{I}_{2\times2} \\
                    \mathbb{I}_{2\times2} & 0 \\
                    \end{array}
                    \right)                   \; ,
\mathbb{J}_{23} =\frac{i}{2}\left(
                    \begin{array}{cc}
                    0 & \sigma^{3} \\
                    -\sigma^{3} & 0 \\
                    \end{array}
                    \right)                   \; , \nonumber
\end{eqnarray}}
where $\mathbb{J}_{03}\equiv \mathbb{P}_{0}$, $\mathbb{J}_{13} \equiv \mathbb{P}_{1}$ and $\mathbb{J}_{23} \equiv \mathbb{P}_{2}$.
The representation \eqref{J_and_P_matrices} shows explicitly that the generators $\mathbb{P}_{a}$ and $\mathbb{J}_{a}$ can be accommodated in the form \eqref{FirstJ} and \eqref{FirstP}. Of course, the form of these generators (diagonal or off-diagonal) depends on our choice the Dirac matrices representation taken in \eqref{gamma_explicit}.

\clearpage
%\section*{References}

\bibliographystyle{apsrev4-1}
\bibliography{libraryTorsionGrande}

\begin{thebibliography}{60}%
\makeatletter
\providecommand \@ifxundefined [1]{%
 \@ifx{#1\undefined}
}%
\providecommand \@ifnum [1]{%
 \ifnum #1\expandafter \@firstoftwo
 \else \expandafter \@secondoftwo
 \fi
}%
\providecommand \@ifx [1]{%
 \ifx #1\expandafter \@firstoftwo
 \else \expandafter \@secondoftwo
 \fi
}%
\providecommand \natexlab [1]{#1}%
\providecommand \enquote  [1]{``#1''}%
\providecommand \bibnamefont  [1]{#1}%
\providecommand \bibfnamefont [1]{#1}%
\providecommand \citenamefont [1]{#1}%
\providecommand \href@noop [0]{\@secondoftwo}%
\providecommand \href [0]{\begingroup \@sanitize@url \@href}%
\providecommand \@href[1]{\@@startlink{#1}\@@href}%
\providecommand \@@href[1]{\endgroup#1\@@endlink}%
\providecommand \@sanitize@url [0]{\catcode `\\12\catcode `\$12\catcode
  `\&12\catcode `\#12\catcode `\^12\catcode `\_12\catcode `\%12\relax}%
\providecommand \@@startlink[1]{}%
\providecommand \@@endlink[0]{}%
\providecommand \url  [0]{\begingroup\@sanitize@url \@url }%
\providecommand \@url [1]{\endgroup\@href {#1}{\urlprefix }}%
\providecommand \urlprefix  [0]{URL }%
\providecommand \Eprint [0]{\href }%
\providecommand \doibase [0]{http://dx.doi.org/}%
\providecommand \selectlanguage [0]{\@gobble}%
\providecommand \bibinfo  [0]{\@secondoftwo}%
\providecommand \bibfield  [0]{\@secondoftwo}%
\providecommand \translation [1]{[#1]}%
\providecommand \BibitemOpen [0]{}%
\providecommand \bibitemStop [0]{}%
\providecommand \bibitemNoStop [0]{.\EOS\space}%
\providecommand \EOS [0]{\spacefactor3000\relax}%
\providecommand \BibitemShut  [1]{\csname bibitem#1\endcsname}%
\let\auto@bib@innerbib\@empty
%</preamble>
\bibitem [{\citenamefont {Semenoff}(1984)}]{semenoff1984}%
  \BibitemOpen
  \bibfield  {author} {\bibinfo {author} {\bibfnamefont {G.~W.}\ \bibnamefont
  {Semenoff}},\ }\href {\doibase 10.1103/PhysRevLett.53.2449} {\bibfield
  {journal} {\bibinfo  {journal} {Phys. Rev. Lett.}\ }\textbf {\bibinfo
  {volume} {53}},\ \bibinfo {pages} {2449} (\bibinfo {year}
  {1984})}\BibitemShut {NoStop}%
\bibitem [{\citenamefont {Gonzalez}\ \emph {et~al.}(1993)\citenamefont
  {Gonzalez}, \citenamefont {Guinea},\ and\ \citenamefont
  {Vozmediano}}]{GONZALEZ1993771}%
  \BibitemOpen
  \bibfield  {author} {\bibinfo {author} {\bibfnamefont {J.}~\bibnamefont
  {Gonzalez}}, \bibinfo {author} {\bibfnamefont {F.}~\bibnamefont {Guinea}}, \
  and\ \bibinfo {author} {\bibfnamefont {M.}~\bibnamefont {Vozmediano}},\
  }\href {\doibase https://doi.org/10.1016/0550-3213(93)90009-E} {\bibfield
  {journal} {\bibinfo  {journal} {Nuclear Physics B}\ }\textbf {\bibinfo
  {volume} {406}},\ \bibinfo {pages} {771 } (\bibinfo {year}
  {1993})}\BibitemShut {NoStop}%
\bibitem [{\citenamefont {Gonz\'alez}\ \emph {et~al.}(1992)\citenamefont
  {Gonz\'alez}, \citenamefont {Guinea},\ and\ \citenamefont
  {Vozmediano}}]{GonzalezFullerenePRL}%
  \BibitemOpen
  \bibfield  {author} {\bibinfo {author} {\bibfnamefont {J.}~\bibnamefont
  {Gonz\'alez}}, \bibinfo {author} {\bibfnamefont {F.}~\bibnamefont {Guinea}},
  \ and\ \bibinfo {author} {\bibfnamefont {M.~A.~H.}\ \bibnamefont
  {Vozmediano}},\ }\href {\doibase 10.1103/PhysRevLett.69.172} {\bibfield
  {journal} {\bibinfo  {journal} {Phys. Rev. Lett.}\ }\textbf {\bibinfo
  {volume} {69}},\ \bibinfo {pages} {172} (\bibinfo {year} {1992})}\BibitemShut
  {NoStop}%
\bibitem [{\citenamefont {Novoselov}\ \emph {et~al.}(2004)\citenamefont
  {Novoselov}, \citenamefont {Geim}, \citenamefont {Morozov}, \citenamefont
  {Jiang}, \citenamefont {Zhang}, \citenamefont {Dubonos}, \citenamefont
  {Grigorieva},\ and\ \citenamefont {Firsov}}]{Novoselov666}%
  \BibitemOpen
  \bibfield  {author} {\bibinfo {author} {\bibfnamefont {K.~S.}\ \bibnamefont
  {Novoselov}}, \bibinfo {author} {\bibfnamefont {A.~K.}\ \bibnamefont {Geim}},
  \bibinfo {author} {\bibfnamefont {S.~V.}\ \bibnamefont {Morozov}}, \bibinfo
  {author} {\bibfnamefont {D.}~\bibnamefont {Jiang}}, \bibinfo {author}
  {\bibfnamefont {Y.}~\bibnamefont {Zhang}}, \bibinfo {author} {\bibfnamefont
  {S.~V.}\ \bibnamefont {Dubonos}}, \bibinfo {author} {\bibfnamefont {I.~V.}\
  \bibnamefont {Grigorieva}}, \ and\ \bibinfo {author} {\bibfnamefont {A.~A.}\
  \bibnamefont {Firsov}},\ }\href {\doibase 10.1126/science.1102896} {\bibfield
   {journal} {\bibinfo  {journal} {Science}\ }\textbf {\bibinfo {volume}
  {306}},\ \bibinfo {pages} {666} (\bibinfo {year} {2004})},\ \Eprint
  {http://arxiv.org/abs/0410550} {arXiv:0410550 [cond-mat]} \BibitemShut
  {NoStop}%
\bibitem [{\citenamefont {Castro~Neto}\ \emph {et~al.}(2009)\citenamefont
  {Castro~Neto}, \citenamefont {Guinea}, \citenamefont {Peres}, \citenamefont
  {Novoselov},\ and\ \citenamefont {Geim}}]{PacoReview2009}%
  \BibitemOpen
  \bibfield  {author} {\bibinfo {author} {\bibfnamefont {A.~H.}\ \bibnamefont
  {Castro~Neto}}, \bibinfo {author} {\bibfnamefont {F.}~\bibnamefont {Guinea}},
  \bibinfo {author} {\bibfnamefont {N.~M.~R.}\ \bibnamefont {Peres}}, \bibinfo
  {author} {\bibfnamefont {K.~S.}\ \bibnamefont {Novoselov}}, \ and\ \bibinfo
  {author} {\bibfnamefont {A.~K.}\ \bibnamefont {Geim}},\ }\href {\doibase
  10.1103/RevModPhys.81.109} {\bibfield  {journal} {\bibinfo  {journal} {Rev.
  Mod. Phys.}\ }\textbf {\bibinfo {volume} {81}},\ \bibinfo {pages} {109}
  (\bibinfo {year} {2009})}\BibitemShut {NoStop}%
\bibitem [{\citenamefont {Yan}\ and\ \citenamefont {Felser}(2017)}]{Claudia}%
  \BibitemOpen
  \bibfield  {author} {\bibinfo {author} {\bibfnamefont {B.}~\bibnamefont
  {Yan}}\ and\ \bibinfo {author} {\bibfnamefont {C.}~\bibnamefont {Felser}},\
  }\href {\doibase 10.1146/annurev-conmatphys-031016-025458} {\bibfield
  {journal} {\bibinfo  {journal} {Annual Review of Condensed Matter Physics}\
  }\textbf {\bibinfo {volume} {8}},\ \bibinfo {pages} {337} (\bibinfo {year}
  {2017})}\BibitemShut {NoStop}%
\bibitem [{\citenamefont {Jakubsk\'y}\ and\ \citenamefont
  {Tu\v{s}ek}(2017)}]{Jakubsky2016}%
  \BibitemOpen
  \bibfield  {author} {\bibinfo {author} {\bibfnamefont {V.}~\bibnamefont
  {Jakubsk\'y}}\ and\ \bibinfo {author} {\bibfnamefont {M.}~\bibnamefont
  {Tu\v{s}ek}},\ }\href {\doibase 10.1016/j.aop.2017.01.016} {\bibfield
  {journal} {\bibinfo  {journal} {Annals Phys.}\ }\textbf {\bibinfo {volume}
  {378}},\ \bibinfo {pages} {171} (\bibinfo {year} {2017})},\ \Eprint
  {http://arxiv.org/abs/1604.00157} {arXiv:1604.00157 [cond-mat.mes-hall]}
  \BibitemShut {NoStop}%
%%CITATION = ARXIV:1604.00157;%%
\bibitem [{\citenamefont {Armitage}\ \emph {et~al.}(2018)\citenamefont
  {Armitage}, \citenamefont {Mele},\ and\ \citenamefont
  {Vishwanath}}]{RevModPhys.90.015001}%
  \BibitemOpen
  \bibfield  {author} {\bibinfo {author} {\bibfnamefont {N.~P.}\ \bibnamefont
  {Armitage}}, \bibinfo {author} {\bibfnamefont {E.~J.}\ \bibnamefont {Mele}},
  \ and\ \bibinfo {author} {\bibfnamefont {A.}~\bibnamefont {Vishwanath}},\
  }\href {\doibase 10.1103/RevModPhys.90.015001} {\bibfield  {journal}
  {\bibinfo  {journal} {Rev. Mod. Phys.}\ }\textbf {\bibinfo {volume} {90}},\
  \bibinfo {pages} {015001} (\bibinfo {year} {2018})}\BibitemShut {NoStop}%
\bibitem [{\citenamefont {Iorio}\ and\ \citenamefont
  {Lambiase}(2014)}]{Iorio:2013ifa}%
  \BibitemOpen
  \bibfield  {author} {\bibinfo {author} {\bibfnamefont {A.}~\bibnamefont
  {Iorio}}\ and\ \bibinfo {author} {\bibfnamefont {G.}~\bibnamefont
  {Lambiase}},\ }\href {\doibase 10.1103/PhysRevD.90.025006} {\bibfield
  {journal} {\bibinfo  {journal} {Phys. Rev.}\ }\textbf {\bibinfo {volume}
  {D90}},\ \bibinfo {pages} {025006} (\bibinfo {year} {2014})},\ \Eprint
  {http://arxiv.org/abs/1308.0265} {arXiv:1308.0265 [hep-th]} \BibitemShut
  {NoStop}%
%%CITATION = ARXIV:1308.0265;%%
\bibitem [{\citenamefont {Iorio}\ and\ \citenamefont
  {Lambiase}(2012)}]{Iorio:2011yz}%
  \BibitemOpen
  \bibfield  {author} {\bibinfo {author} {\bibfnamefont {A.}~\bibnamefont
  {Iorio}}\ and\ \bibinfo {author} {\bibfnamefont {G.}~\bibnamefont
  {Lambiase}},\ }\href {\doibase 10.1016/j.physletb.2012.08.023} {\bibfield
  {journal} {\bibinfo  {journal} {Phys. Lett.}\ }\textbf {\bibinfo {volume}
  {B716}},\ \bibinfo {pages} {334} (\bibinfo {year} {2012})},\ \Eprint
  {http://arxiv.org/abs/1108.2340} {arXiv:1108.2340 [cond-mat.mtrl-sci]}
  \BibitemShut {NoStop}%
%%CITATION = ARXIV:1108.2340;%%
\bibitem [{\citenamefont {Iorio}\ and\ \citenamefont {Pais}(2015)}]{IorioPais}%
  \BibitemOpen
  \bibfield  {author} {\bibinfo {author} {\bibfnamefont {A.}~\bibnamefont
  {Iorio}}\ and\ \bibinfo {author} {\bibfnamefont {P.}~\bibnamefont {Pais}},\
  }\href {\doibase 10.1103/PhysRevD.92.125005} {\bibfield  {journal} {\bibinfo
  {journal} {Phys. Rev. D}\ }\textbf {\bibinfo {volume} {92}},\ \bibinfo
  {pages} {125005} (\bibinfo {year} {2015})},\ \Eprint
  {http://arxiv.org/abs/1508.00926} {arXiv:1508.00926 [hep-th]} \BibitemShut
  {NoStop}%
%%CITATION = ARXIV:1508.00926;%%
\bibitem [{\citenamefont {Iorio}(2015{\natexlab{a}})}]{IorioReview}%
  \BibitemOpen
  \bibfield  {author} {\bibinfo {author} {\bibfnamefont {A.}~\bibnamefont
  {Iorio}},\ }\href {\doibase 10.1142/S021827181530013X} {\bibfield  {journal}
  {\bibinfo  {journal} {Int. J. Mod. Phys.}\ }\textbf {\bibinfo {volume}
  {D24}},\ \bibinfo {pages} {1530013} (\bibinfo {year} {2015}{\natexlab{a}})},\
  \Eprint {http://arxiv.org/abs/1412.4554} {arXiv:1412.4554 [hep-th]}
  \BibitemShut {NoStop}%
%%CITATION = ARXIV:1412.4554;%%
\bibitem [{\citenamefont {Iorio}\ \emph {et~al.}(2018)\citenamefont {Iorio},
  \citenamefont {Pais}, \citenamefont {Elmashad}, \citenamefont {Ali},
  \citenamefont {Faizal},\ and\ \citenamefont {Abou-Salem}}]{Iorio:2017vtw}%
  \BibitemOpen
  \bibfield  {author} {\bibinfo {author} {\bibfnamefont {A.}~\bibnamefont
  {Iorio}}, \bibinfo {author} {\bibfnamefont {P.}~\bibnamefont {Pais}},
  \bibinfo {author} {\bibfnamefont {I.~A.}\ \bibnamefont {Elmashad}}, \bibinfo
  {author} {\bibfnamefont {A.~F.}\ \bibnamefont {Ali}}, \bibinfo {author}
  {\bibfnamefont {M.}~\bibnamefont {Faizal}}, \ and\ \bibinfo {author}
  {\bibfnamefont {L.~I.}\ \bibnamefont {Abou-Salem}},\ }\href {\doibase
  10.1142/S0218271818500803} {\bibfield  {journal} {\bibinfo  {journal} {Int.
  J. Mod. Phys.}\ }\textbf {\bibinfo {volume} {D27}},\ \bibinfo {pages}
  {1850080} (\bibinfo {year} {2018})},\ \Eprint
  {http://arxiv.org/abs/1706.01332} {arXiv:1706.01332 [physics.gen-ph]}
  \BibitemShut {NoStop}%
%%CITATION = ARXIV:1706.01332;%%
\bibitem [{\citenamefont {Acquaviva}\ \emph
  {et~al.}(2017{\natexlab{a}})\citenamefont {Acquaviva}, \citenamefont
  {Iorio},\ and\ \citenamefont {Scholtz}}]{Acquaviva:2017xqi}%
  \BibitemOpen
  \bibfield  {author} {\bibinfo {author} {\bibfnamefont {G.}~\bibnamefont
  {Acquaviva}}, \bibinfo {author} {\bibfnamefont {A.}~\bibnamefont {Iorio}}, \
  and\ \bibinfo {author} {\bibfnamefont {M.}~\bibnamefont {Scholtz}},\ }\href
  {\doibase 10.1016/j.aop.2017.10.018} {\bibfield  {journal} {\bibinfo
  {journal} {Annals Phys.}\ }\textbf {\bibinfo {volume} {387}},\ \bibinfo
  {pages} {317} (\bibinfo {year} {2017}{\natexlab{a}})},\ \Eprint
  {http://arxiv.org/abs/1704.00345} {arXiv:1704.00345 [gr-qc]} \BibitemShut
  {NoStop}%
%%CITATION = ARXIV:1704.00345;%%
\bibitem [{\citenamefont {Acquaviva}\ \emph
  {et~al.}(2017{\natexlab{b}})\citenamefont {Acquaviva}, \citenamefont
  {Iorio},\ and\ \citenamefont {Scholtz}}]{Acquaviva:2017krr}%
  \BibitemOpen
  \bibfield  {author} {\bibinfo {author} {\bibfnamefont {G.}~\bibnamefont
  {Acquaviva}}, \bibinfo {author} {\bibfnamefont {A.}~\bibnamefont {Iorio}}, \
  and\ \bibinfo {author} {\bibfnamefont {M.}~\bibnamefont {Scholtz}},\ }in\
  \href {https://inspirehep.net/record/1643322/files/arXiv:1712.05275.pdf}
  {\emph {\bibinfo {booktitle} {{17th Hellenic School and Workshops on
  Elementary Particle Physics and Gravity (CORFU2017) Corfu, Greece, September
  2-28, 2017}}}}\ (\bibinfo {year} {2017})\ \Eprint
  {http://arxiv.org/abs/1712.05275} {arXiv:1712.05275 [hep-th]} \BibitemShut
  {NoStop}%
%%CITATION = ARXIV:1712.05275;%%
\bibitem [{\citenamefont {{Dardashti}}\ \emph {et~al.}(2016)\citenamefont
  {{Dardashti}}, \citenamefont {{Hartmann}}, \citenamefont {{Th{\'e}bault}},\
  and\ \citenamefont {{Winsberg}}}]{Dardashti2016arXiv160405932D}%
  \BibitemOpen
  \bibfield  {author} {\bibinfo {author} {\bibfnamefont {R.}~\bibnamefont
  {{Dardashti}}}, \bibinfo {author} {\bibfnamefont {S.}~\bibnamefont
  {{Hartmann}}}, \bibinfo {author} {\bibfnamefont {K.~P.~Y.}\ \bibnamefont
  {{Th{\'e}bault}}}, \ and\ \bibinfo {author} {\bibfnamefont {E.}~\bibnamefont
  {{Winsberg}}},\ }\href@noop {} {\bibfield  {journal} {\bibinfo  {journal}
  {ArXiv e-prints}\ } (\bibinfo {year} {2016})},\ \Eprint
  {http://arxiv.org/abs/1604.05932} {arXiv:1604.05932} \BibitemShut {NoStop}%
\bibitem [{\citenamefont {Volovik}(2006)}]{Volovik:2003fe}%
  \BibitemOpen
  \bibfield  {author} {\bibinfo {author} {\bibfnamefont {G.~E.}\ \bibnamefont
  {Volovik}},\ }\href@noop {} {\bibfield  {journal} {\bibinfo  {journal} {Int.
  Ser. Monogr. Phys.}\ }\textbf {\bibinfo {volume} {117}},\ \bibinfo {pages}
  {1} (\bibinfo {year} {2006})}\BibitemShut {NoStop}%
%%CITATION = IMPHA,117,1;%%
\bibitem [{\citenamefont {Barcel{\'o}}\ \emph {et~al.}(2011)\citenamefont
  {Barcel{\'o}}, \citenamefont {Liberati},\ and\ \citenamefont
  {Visser}}]{Barceló2011}%
  \BibitemOpen
  \bibfield  {author} {\bibinfo {author} {\bibfnamefont {C.}~\bibnamefont
  {Barcel{\'o}}}, \bibinfo {author} {\bibfnamefont {S.}~\bibnamefont
  {Liberati}}, \ and\ \bibinfo {author} {\bibfnamefont {M.}~\bibnamefont
  {Visser}},\ }\href {\doibase 10.12942/lrr-2011-3} {\bibfield  {journal}
  {\bibinfo  {journal} {Living Reviews in Relativity}\ }\textbf {\bibinfo
  {volume} {14}},\ \bibinfo {pages} {3} (\bibinfo {year} {2011})}\BibitemShut
  {NoStop}%
\bibitem [{\citenamefont {Steinhauer}(2016)}]{Steinhauer:2015saa}%
  \BibitemOpen
  \bibfield  {author} {\bibinfo {author} {\bibfnamefont {J.}~\bibnamefont
  {Steinhauer}},\ }\href {\doibase 10.1038/nphys3863} {\bibfield  {journal}
  {\bibinfo  {journal} {Nature Phys.}\ }\textbf {\bibinfo {volume} {12}},\
  \bibinfo {pages} {959} (\bibinfo {year} {2016})},\ \Eprint
  {http://arxiv.org/abs/1510.00621} {arXiv:1510.00621 [gr-qc]} \BibitemShut
  {NoStop}%
%%CITATION = ARXIV:1510.00621;%%
\bibitem [{\citenamefont {Castorina}\ \emph {et~al.}(2007)\citenamefont
  {Castorina}, \citenamefont {Kharzeev},\ and\ \citenamefont
  {Satz}}]{Castorina:2007eb}%
  \BibitemOpen
  \bibfield  {author} {\bibinfo {author} {\bibfnamefont {P.}~\bibnamefont
  {Castorina}}, \bibinfo {author} {\bibfnamefont {D.}~\bibnamefont {Kharzeev}},
  \ and\ \bibinfo {author} {\bibfnamefont {H.}~\bibnamefont {Satz}},\ }\href
  {\doibase 10.1140/epjc/s10052-007-0368-6} {\bibfield  {journal} {\bibinfo
  {journal} {Eur. Phys. J.}\ }\textbf {\bibinfo {volume} {C52}},\ \bibinfo
  {pages} {187} (\bibinfo {year} {2007})},\ \Eprint
  {http://arxiv.org/abs/0704.1426} {arXiv:0704.1426 [hep-ph]} \BibitemShut
  {NoStop}%
%%CITATION = ARXIV:0704.1426;%%
\bibitem [{\citenamefont {Castorina}\ \emph {et~al.}(2008)\citenamefont
  {Castorina}, \citenamefont {Grumiller},\ and\ \citenamefont
  {Iorio}}]{Castorina:2008gf}%
  \BibitemOpen
  \bibfield  {author} {\bibinfo {author} {\bibfnamefont {P.}~\bibnamefont
  {Castorina}}, \bibinfo {author} {\bibfnamefont {D.}~\bibnamefont
  {Grumiller}}, \ and\ \bibinfo {author} {\bibfnamefont {A.}~\bibnamefont
  {Iorio}},\ }\href {\doibase 10.1103/PhysRevD.77.124034} {\bibfield  {journal}
  {\bibinfo  {journal} {Phys. Rev.}\ }\textbf {\bibinfo {volume} {D77}},\
  \bibinfo {pages} {124034} (\bibinfo {year} {2008})},\ \Eprint
  {http://arxiv.org/abs/0802.2286} {arXiv:0802.2286 [hep-th]} \BibitemShut
  {NoStop}%
%%CITATION = ARXIV:0802.2286;%%
\bibitem [{\citenamefont {Castorina}\ \emph {et~al.}(2015)\citenamefont
  {Castorina}, \citenamefont {Iorio},\ and\ \citenamefont
  {Satz}}]{Castorina:2014fna}%
  \BibitemOpen
  \bibfield  {author} {\bibinfo {author} {\bibfnamefont {P.}~\bibnamefont
  {Castorina}}, \bibinfo {author} {\bibfnamefont {A.}~\bibnamefont {Iorio}}, \
  and\ \bibinfo {author} {\bibfnamefont {H.}~\bibnamefont {Satz}},\ }\href
  {\doibase 10.1142/S0218301315500561} {\bibfield  {journal} {\bibinfo
  {journal} {Int. J. Mod. Phys.}\ }\textbf {\bibinfo {volume} {E24}},\ \bibinfo
  {pages} {1550056} (\bibinfo {year} {2015})},\ \Eprint
  {http://arxiv.org/abs/1409.3104} {arXiv:1409.3104 [hep-ph]} \BibitemShut
  {NoStop}%
%%CITATION = ARXIV:1409.3104;%%
\bibitem [{\citenamefont {Gooth}\ \emph {et~al.}(2017)\citenamefont {Gooth}
  \emph {et~al.}}]{Gooth:2017mbd}%
  \BibitemOpen
  \bibfield  {author} {\bibinfo {author} {\bibfnamefont {J.}~\bibnamefont
  {Gooth}} \emph {et~al.},\ }\href {\doibase 10.1038/nature23005} {\bibfield
  {journal} {\bibinfo  {journal} {Nature}\ }\textbf {\bibinfo {volume} {547}},\
  \bibinfo {pages} {324} (\bibinfo {year} {2017})},\ \Eprint
  {http://arxiv.org/abs/1703.10682} {arXiv:1703.10682 [cond-mat.mtrl-sci]}
  \BibitemShut {NoStop}%
%%CITATION = ARXIV:1703.10682;%%
\bibitem [{\citenamefont {Dudal}\ \emph {et~al.}(2018)\citenamefont {Dudal},
  \citenamefont {Mizher},\ and\ \citenamefont {Pais}}]{Dudal:2018mms}%
  \BibitemOpen
  \bibfield  {author} {\bibinfo {author} {\bibfnamefont {D.}~\bibnamefont
  {Dudal}}, \bibinfo {author} {\bibfnamefont {A.~J.}\ \bibnamefont {Mizher}}, \
  and\ \bibinfo {author} {\bibfnamefont {P.}~\bibnamefont {Pais}},\ }\href@noop
  {} {\  (\bibinfo {year} {2018})},\ \Eprint {http://arxiv.org/abs/1801.08853}
  {arXiv:1801.08853 [hep-th]} \BibitemShut {NoStop}%
%%CITATION = ARXIV:1801.08853
\bibitem [{\citenamefont {Alvarez}\ \emph {et~al.}(2012)\citenamefont
  {Alvarez}, \citenamefont {Valenzuela},\ and\ \citenamefont {Zanelli}}]{AVZ}%
  \BibitemOpen
  \bibfield  {author} {\bibinfo {author} {\bibfnamefont {P.~D.}\ \bibnamefont
  {Alvarez}}, \bibinfo {author} {\bibfnamefont {M.}~\bibnamefont {Valenzuela}},
  \ and\ \bibinfo {author} {\bibfnamefont {J.}~\bibnamefont {Zanelli}},\ }\href
  {\doibase 10.1007/JHEP04(2012)058} {\bibfield  {journal} {\bibinfo  {journal}
  {JHEP}\ }\textbf {\bibinfo {volume} {1204}},\ \bibinfo {pages} {058}
  (\bibinfo {year} {2012})},\ \Eprint {http://arxiv.org/abs/1109.3944}
  {arXiv:1109.3944 [hep-th]} \BibitemShut {NoStop}%
%%CITATION = ARXIV:1109.3944;%%
\bibitem [{\citenamefont {Alvarez}\ \emph {et~al.}(2015)\citenamefont
  {Alvarez}, \citenamefont {Pais}, \citenamefont {Rodríguez}, \citenamefont
  {Salgado-Rebolledo},\ and\ \citenamefont {Zanelli}}]{u-susy-su2}%
  \BibitemOpen
  \bibfield  {author} {\bibinfo {author} {\bibfnamefont {P.~D.}\ \bibnamefont
  {Alvarez}}, \bibinfo {author} {\bibfnamefont {P.}~\bibnamefont {Pais}},
  \bibinfo {author} {\bibfnamefont {E.}~\bibnamefont {Rodríguez}}, \bibinfo
  {author} {\bibfnamefont {P.}~\bibnamefont {Salgado-Rebolledo}}, \ and\
  \bibinfo {author} {\bibfnamefont {J.}~\bibnamefont {Zanelli}},\ }\href
  {\doibase 10.1088/0264-9381/32/17/175014} {\bibfield  {journal} {\bibinfo
  {journal} {Class. Quant. Grav.}\ }\textbf {\bibinfo {volume} {32}},\ \bibinfo
  {pages} {175014} (\bibinfo {year} {2015})},\ \Eprint
  {http://arxiv.org/abs/1505.03834} {arXiv:1505.03834 [hep-th]} \BibitemShut
  {NoStop}%
%%CITATION = ARXIV:1505.03834;%%
\bibitem [{\citenamefont {Andrianopoli}\ \emph {et~al.}(2018)\citenamefont
  {Andrianopoli}, \citenamefont {Cerchiai}, \citenamefont {D.'Auria},\ and\
  \citenamefont {Trigiante}}]{Andrianopoli2018}%
  \BibitemOpen
  \bibfield  {author} {\bibinfo {author} {\bibfnamefont {L.}~\bibnamefont
  {Andrianopoli}}, \bibinfo {author} {\bibfnamefont {B.~L.}\ \bibnamefont
  {Cerchiai}}, \bibinfo {author} {\bibfnamefont {R.}~\bibnamefont {D.'Auria}},
  \ and\ \bibinfo {author} {\bibfnamefont {M.}~\bibnamefont {Trigiante}},\
  }\href {\doibase 10.1007/JHEP04(2018)007} {\bibfield  {journal} {\bibinfo
  {journal} {JHEP}\ }\textbf {\bibinfo {volume} {04}},\ \bibinfo {pages} {007}
  (\bibinfo {year} {2018})},\ \Eprint {http://arxiv.org/abs/1801.08081}
  {arXiv:1801.08081 [hep-th]} \BibitemShut {NoStop}%
%%CITATION = ARXIV:1801.08081;%%
\bibitem [{\citenamefont {Kleinert}(1989)}]{Kleinert:1989ky}%
  \BibitemOpen
  \bibfield  {author} {\bibinfo {author} {\bibfnamefont {H.}~\bibnamefont
  {Kleinert}},\ }\href@noop {} {\emph {\bibinfo {title} {{Gauge fields in
  condensed matter. Vol. 2: Stresses and defects. Differential geometry,
  crystal melting}}}}\ (\bibinfo {year} {1989})\BibitemShut {NoStop}%
%%CITATION = INSPIRE-289365;%%
\bibitem [{\citenamefont {Katanaev}\ and\ \citenamefont
  {Volovich}(1992)}]{Katanaev:1992kh}%
  \BibitemOpen
  \bibfield  {author} {\bibinfo {author} {\bibfnamefont {M.~O.}\ \bibnamefont
  {Katanaev}}\ and\ \bibinfo {author} {\bibfnamefont {I.~V.}\ \bibnamefont
  {Volovich}},\ }\href {\doibase 10.1016/0003-4916(52)90040-7} {\bibfield
  {journal} {\bibinfo  {journal} {Annals Phys.}\ }\textbf {\bibinfo {volume}
  {216}},\ \bibinfo {pages} {1} (\bibinfo {year} {1992})}\BibitemShut {NoStop}%
%%CITATION = APNYA,216,1;%%
\bibitem [{\citenamefont {Yazyev}\ and\ \citenamefont
  {Louie}(2010)}]{Yazyev2010}%
  \BibitemOpen
  \bibfield  {author} {\bibinfo {author} {\bibfnamefont {O.~V.}\ \bibnamefont
  {Yazyev}}\ and\ \bibinfo {author} {\bibfnamefont {S.~G.}\ \bibnamefont
  {Louie}},\ }\href@noop {} {\bibfield  {journal} {\bibinfo  {journal} {Phys.
  Rev. B}\ }\textbf {\bibinfo {volume} {81}},\ \bibinfo {pages} {195420}
  (\bibinfo {year} {2010})},\ \Eprint {http://arxiv.org/abs/1004.2031}
  {arXiv:1004.2031 [cond-mat]} \BibitemShut {NoStop}%
\bibitem [{\citenamefont {Hirth}\ and\ \citenamefont
  {Lothe}(1967)}]{hirth1967theory}%
  \BibitemOpen
  \bibfield  {author} {\bibinfo {author} {\bibfnamefont {J.}~\bibnamefont
  {Hirth}}\ and\ \bibinfo {author} {\bibfnamefont {J.}~\bibnamefont {Lothe}},\
  }\href {https://books.google.cz/books?id=OgfwAAAAMAAJ} {\emph {\bibinfo
  {title} {Theory of Dislocations}}},\ McGraw-Hill series in electrical
  engineering: Electronics and electronic circuits\ (\bibinfo  {publisher}
  {McGraw-Hill},\ \bibinfo {year} {1967})\BibitemShut {NoStop}%
\bibitem [{\citenamefont {Zhang}\ \emph {et~al.}(2015)\citenamefont {Zhang},
  \citenamefont {Xu}, \citenamefont {Yuan}, \citenamefont {Xin},\ and\
  \citenamefont {Ding}}]{C5NR04960A}%
  \BibitemOpen
  \bibfield  {author} {\bibinfo {author} {\bibfnamefont {X.}~\bibnamefont
  {Zhang}}, \bibinfo {author} {\bibfnamefont {Z.}~\bibnamefont {Xu}}, \bibinfo
  {author} {\bibfnamefont {Q.}~\bibnamefont {Yuan}}, \bibinfo {author}
  {\bibfnamefont {J.}~\bibnamefont {Xin}}, \ and\ \bibinfo {author}
  {\bibfnamefont {F.}~\bibnamefont {Ding}},\ }\href {\doibase
  10.1039/C5NR04960A} {\bibfield  {journal} {\bibinfo  {journal} {Nanoscale}\
  }\textbf {\bibinfo {volume} {7}},\ \bibinfo {pages} {20082} (\bibinfo {year}
  {2015})}\BibitemShut {NoStop}%
\bibitem [{\citenamefont {Katanaev}(2005)}]{Katanaev2005}%
  \BibitemOpen
  \bibfield  {author} {\bibinfo {author} {\bibfnamefont {M.~O.}\ \bibnamefont
  {Katanaev}},\ }\bibfield  {booktitle} {\emph {\bibinfo {booktitle} {{Summer
  School on Vortices: A Unifying Concept in Physics Cargese, Corsica, France,
  July 5-16, 2004}}},\ }\href {\doibase 10.1070/PU2005v048n07ABEH002027}
  {\bibfield  {journal} {\bibinfo  {journal} {Phys. Usp.}\ }\textbf {\bibinfo
  {volume} {48}},\ \bibinfo {pages} {675} (\bibinfo {year} {2005})},\ \bibinfo
  {note} {[Usp. Fiz. Nauk175,705(2005)]},\ \Eprint
  {http://arxiv.org/abs/cond-mat/0407469} {arXiv:cond-mat/0407469
  [cond-mat.mtrl-sci]} \BibitemShut {NoStop}%
%%CITATION = COND-MAT/0407469;%%
\bibitem [{\citenamefont {Deser}\ \emph
  {et~al.}(1982{\natexlab{a}})\citenamefont {Deser}, \citenamefont {Jackiw},\
  and\ \citenamefont {Templeton}}]{Deser:1981wh}%
  \BibitemOpen
  \bibfield  {author} {\bibinfo {author} {\bibfnamefont {S.}~\bibnamefont
  {Deser}}, \bibinfo {author} {\bibfnamefont {R.}~\bibnamefont {Jackiw}}, \
  and\ \bibinfo {author} {\bibfnamefont {S.}~\bibnamefont {Templeton}},\ }\href
  {\doibase 10.1006/aphy.2000.6013, 10.1016/0003-4916(82)90164-6} {\bibfield
  {journal} {\bibinfo  {journal} {Annals Phys.}\ }\textbf {\bibinfo {volume}
  {140}},\ \bibinfo {pages} {372} (\bibinfo {year} {1982}{\natexlab{a}})},\
  \bibinfo {note} {[Annals Phys.281,409(2000)]}\BibitemShut {NoStop}%
%%CITATION = APNYA,140,372;%%
\bibitem [{\citenamefont {Deser}\ \emph
  {et~al.}(1982{\natexlab{b}})\citenamefont {Deser}, \citenamefont {Jackiw},\
  and\ \citenamefont {Templeton}}]{Deser:1982vy}%
  \BibitemOpen
  \bibfield  {author} {\bibinfo {author} {\bibfnamefont {S.}~\bibnamefont
  {Deser}}, \bibinfo {author} {\bibfnamefont {R.}~\bibnamefont {Jackiw}}, \
  and\ \bibinfo {author} {\bibfnamefont {S.}~\bibnamefont {Templeton}},\ }\href
  {\doibase 10.1103/PhysRevLett.48.975} {\bibfield  {journal} {\bibinfo
  {journal} {Phys. Rev. Lett.}\ }\textbf {\bibinfo {volume} {48}},\ \bibinfo
  {pages} {975} (\bibinfo {year} {1982}{\natexlab{b}})}\BibitemShut {NoStop}%
%%CITATION = PRLTA,48,975;%%
\bibitem [{\citenamefont {Horne}\ and\ \citenamefont
  {Witten}(1989)}]{Horne:1988jf}%
  \BibitemOpen
  \bibfield  {author} {\bibinfo {author} {\bibfnamefont {J.~H.}\ \bibnamefont
  {Horne}}\ and\ \bibinfo {author} {\bibfnamefont {E.}~\bibnamefont {Witten}},\
  }\href {\doibase 10.1103/PhysRevLett.62.501} {\bibfield  {journal} {\bibinfo
  {journal} {Phys. Rev. Lett.}\ }\textbf {\bibinfo {volume} {62}},\ \bibinfo
  {pages} {501} (\bibinfo {year} {1989})}\BibitemShut {NoStop}%
%%CITATION = PRLTA,62,501;%%
\bibitem [{\citenamefont {Guralnik}\ \emph {et~al.}(2003)\citenamefont
  {Guralnik}, \citenamefont {Iorio}, \citenamefont {Jackiw},\ and\
  \citenamefont {Pi}}]{Guralnik:2003we}%
  \BibitemOpen
  \bibfield  {author} {\bibinfo {author} {\bibfnamefont {G.}~\bibnamefont
  {Guralnik}}, \bibinfo {author} {\bibfnamefont {A.}~\bibnamefont {Iorio}},
  \bibinfo {author} {\bibfnamefont {R.}~\bibnamefont {Jackiw}}, \ and\ \bibinfo
  {author} {\bibfnamefont {S.~Y.}\ \bibnamefont {Pi}},\ }\href {\doibase
  10.1016/S0003-4916(03)00142-8} {\bibfield  {journal} {\bibinfo  {journal}
  {Annals Phys.}\ }\textbf {\bibinfo {volume} {308}},\ \bibinfo {pages} {222}
  (\bibinfo {year} {2003})},\ \Eprint {http://arxiv.org/abs/hep-th/0305117}
  {arXiv:hep-th/0305117 [hep-th]} \BibitemShut {NoStop}%
%%CITATION = HEP-TH/0305117;%%
\bibitem [{\citenamefont {Witten}(1988)}]{Witten1988}%
  \BibitemOpen
  \bibfield  {author} {\bibinfo {author} {\bibfnamefont {E.}~\bibnamefont
  {Witten}},\ }\href {\doibase 10.1016/0550-3213(88)90143-5} {\bibfield
  {journal} {\bibinfo  {journal} {Nucl. Phys.}\ }\textbf {\bibinfo {volume}
  {B311}},\ \bibinfo {pages} {46} (\bibinfo {year} {1988})}\BibitemShut
  {NoStop}%
%%CITATION = NUPHA,B311,46;%%
\bibitem [{\citenamefont {Yazyev}\ and\ \citenamefont
  {Chen}(2014)}]{Yazyev2014}%
  \BibitemOpen
  \bibfield  {author} {\bibinfo {author} {\bibfnamefont {O.~V.}\ \bibnamefont
  {Yazyev}}\ and\ \bibinfo {author} {\bibfnamefont {Y.~P.}\ \bibnamefont
  {Chen}},\ }\href@noop {} {\bibfield  {journal} {\bibinfo  {journal} {Nature
  nanotechnology}\ }\textbf {\bibinfo {volume} {9}},\ \bibinfo {pages} {755}
  (\bibinfo {year} {2014})},\ \Eprint {http://arxiv.org/abs/1502.04899}
  {arXiv:1502.04899 [cond-mat]} \BibitemShut {NoStop}%
\bibitem [{\citenamefont {Iorio}(2015{\natexlab{b}})}]{Iorio:2015iha}%
  \BibitemOpen
  \bibfield  {author} {\bibinfo {author} {\bibfnamefont {A.}~\bibnamefont
  {Iorio}},\ }\bibfield  {booktitle} {\emph {\bibinfo {booktitle}
  {{Proceedings, 7th International Workshop : Spacetime - Matter - Quantum
  Mechanics. (DICE2014): Castiglioncello, Tuscany, Italy, September 15-19,
  2014}}},\ }\href {\doibase 10.1088/1742-6596/626/1/012035} {\bibfield
  {journal} {\bibinfo  {journal} {J. Phys. Conf. Ser.}\ }\textbf {\bibinfo
  {volume} {626}},\ \bibinfo {pages} {012035} (\bibinfo {year}
  {2015}{\natexlab{b}})}\BibitemShut {NoStop}%
%%CITATION = 00462,626,012035;%%
\bibitem [{\citenamefont {Nakahara}(2003)}]{Nakahara}%
  \BibitemOpen
  \bibfield  {author} {\bibinfo {author} {\bibfnamefont {M.}~\bibnamefont
  {Nakahara}},\ }\href {https://books.google.fr/books?id=cH-XQB0Ex5wC} {\emph
  {\bibinfo {title} {Geometry, Topology and Physics, Second Edition}}},\
  Graduate student series in physics\ (\bibinfo  {publisher} {Taylor \&
  Francis},\ \bibinfo {year} {2003})\BibitemShut {NoStop}%
\bibitem [{\citenamefont {Goerbig}(2011)}]{RevModPhys.83.1193}%
  \BibitemOpen
  \bibfield  {author} {\bibinfo {author} {\bibfnamefont {M.~O.}\ \bibnamefont
  {Goerbig}},\ }\href {\doibase 10.1103/RevModPhys.83.1193} {\bibfield
  {journal} {\bibinfo  {journal} {Rev. Mod. Phys.}\ }\textbf {\bibinfo {volume}
  {83}},\ \bibinfo {pages} {1193} (\bibinfo {year} {2011})}\BibitemShut
  {NoStop}%
\bibitem [{\citenamefont {Hassaine}\ and\ \citenamefont
  {Zanelli}(2016)}]{Z-book}%
  \BibitemOpen
  \bibfield  {author} {\bibinfo {author} {\bibfnamefont {M.}~\bibnamefont
  {Hassaine}}\ and\ \bibinfo {author} {\bibfnamefont {J.}~\bibnamefont
  {Zanelli}},\ }\href {https://books.google.es/books?id=vjOFjgEACAAJ} {\emph
  {\bibinfo {title} {Chern-Simons (Super)Gravity}}},\ 100 Years of General
  Relativity\ (\bibinfo  {publisher} {World Scientific Publishing Company Pte
  Limited},\ \bibinfo {year} {2016})\BibitemShut {NoStop}%
\bibitem [{\citenamefont {Gilmore}(2012)}]{Gilmore-book}%
  \BibitemOpen
  \bibfield  {author} {\bibinfo {author} {\bibfnamefont {R.}~\bibnamefont
  {Gilmore}},\ }\href {https://books.google.es/books?id=ePMX38H4DD4C} {\emph
  {\bibinfo {title} {Lie Groups, Lie Algebras, and Some of Their
  Applications}}},\ Dover Books on Mathematics\ (\bibinfo  {publisher} {Dover
  Publications},\ \bibinfo {year} {2012})\BibitemShut {NoStop}%
\bibitem [{\citenamefont {Peskin}\ and\ \citenamefont
  {Schroeder}(1995)}]{Peskin}%
  \BibitemOpen
  \bibfield  {author} {\bibinfo {author} {\bibfnamefont {M.}~\bibnamefont
  {Peskin}}\ and\ \bibinfo {author} {\bibfnamefont {D.}~\bibnamefont
  {Schroeder}},\ }\href@noop {} {\emph {\bibinfo {title} {An Introduction to
  Quantum Field Theory}}},\ Advanced book classics\ (\bibinfo  {publisher}
  {Addison-Wesley Publishing Company},\ \bibinfo {year} {1995})\BibitemShut
  {NoStop}%
\bibitem [{\citenamefont {Shapiro}(2002)}]{Shapiro}%
  \BibitemOpen
  \bibfield  {author} {\bibinfo {author} {\bibfnamefont {I.~L.}\ \bibnamefont
  {Shapiro}},\ }\href {\doibase 10.1016/S0370-1573(01)00030-8} {\bibfield
  {journal} {\bibinfo  {journal} {Phys. Rept.}\ }\textbf {\bibinfo {volume}
  {357}},\ \bibinfo {pages} {113} (\bibinfo {year} {2002})},\ \Eprint
  {http://arxiv.org/abs/hep-th/0103093} {arXiv:hep-th/0103093 [hep-th]}
  \BibitemShut {NoStop}%
%%CITATION = HEP-TH/0103093;%%
\bibitem [{\citenamefont {Gusynin}\ \emph {et~al.}(2007)\citenamefont
  {Gusynin}, \citenamefont {Sharapov},\ and\ \citenamefont
  {Carbotte}}]{Gusynin2007}%
  \BibitemOpen
  \bibfield  {author} {\bibinfo {author} {\bibfnamefont {V.~P.}\ \bibnamefont
  {Gusynin}}, \bibinfo {author} {\bibfnamefont {S.~G.}\ \bibnamefont
  {Sharapov}}, \ and\ \bibinfo {author} {\bibfnamefont {J.~P.}\ \bibnamefont
  {Carbotte}},\ }\href {\doibase 10.1142/S0217979207038022} {\bibfield
  {journal} {\bibinfo  {journal} {Int. J. Mod. Phys.}\ }\textbf {\bibinfo
  {volume} {B21}},\ \bibinfo {pages} {4611} (\bibinfo {year} {2007})},\ \Eprint
  {http://arxiv.org/abs/0706.3016} {arXiv:0706.3016 [cond-mat]} \BibitemShut
  {NoStop}%
%%CITATION = ARXIV:0706.3016;%%
\bibitem [{\citenamefont {Banados}\ \emph {et~al.}(1992)\citenamefont
  {Banados}, \citenamefont {Teitelboim},\ and\ \citenamefont
  {Zanelli}}]{BTZ1992}%
  \BibitemOpen
  \bibfield  {author} {\bibinfo {author} {\bibfnamefont {M.}~\bibnamefont
  {Banados}}, \bibinfo {author} {\bibfnamefont {C.}~\bibnamefont {Teitelboim}},
  \ and\ \bibinfo {author} {\bibfnamefont {J.}~\bibnamefont {Zanelli}},\ }\href
  {\doibase 10.1103/PhysRevLett.69.1849} {\bibfield  {journal} {\bibinfo
  {journal} {Phys. Rev. Lett.}\ }\textbf {\bibinfo {volume} {69}},\ \bibinfo
  {pages} {1849} (\bibinfo {year} {1992})},\ \Eprint
  {http://arxiv.org/abs/hep-th/9204099} {arXiv:hep-th/9204099 [hep-th]}
  \BibitemShut {NoStop}%
%%CITATION = HEP-TH/9204099;%%
\bibitem [{\citenamefont {Guevara}\ \emph {et~al.}(2016)\citenamefont
  {Guevara}, \citenamefont {Pais},\ and\ \citenamefont {Zanelli}}]{GPZ}%
  \BibitemOpen
  \bibfield  {author} {\bibinfo {author} {\bibfnamefont {A.}~\bibnamefont
  {Guevara}}, \bibinfo {author} {\bibfnamefont {P.}~\bibnamefont {Pais}}, \
  and\ \bibinfo {author} {\bibfnamefont {J.}~\bibnamefont {Zanelli}},\ }\href
  {\doibase 10.1007/JHEP08(2016)085} {\bibfield  {journal} {\bibinfo  {journal}
  {JHEP}\ }\textbf {\bibinfo {volume} {08}},\ \bibinfo {pages} {085} (\bibinfo
  {year} {2016})},\ \Eprint {http://arxiv.org/abs/1606.05239} {arXiv:1606.05239
  [hep-th]} \BibitemShut {NoStop}%
%%CITATION = ARXIV:1606.05239;%%
\bibitem [{\citenamefont {Mesaros}\ \emph {et~al.}(2010)\citenamefont
  {Mesaros}, \citenamefont {Sadri},\ and\ \citenamefont
  {Zaanen}}]{Mesaros:2009az}%
  \BibitemOpen
  \bibfield  {author} {\bibinfo {author} {\bibfnamefont {A.}~\bibnamefont
  {Mesaros}}, \bibinfo {author} {\bibfnamefont {D.}~\bibnamefont {Sadri}}, \
  and\ \bibinfo {author} {\bibfnamefont {J.}~\bibnamefont {Zaanen}},\ }\href
  {\doibase 10.1103/PhysRevB.82.073405} {\bibfield  {journal} {\bibinfo
  {journal} {Phys. Rev.}\ }\textbf {\bibinfo {volume} {B82}},\ \bibinfo {pages}
  {073405} (\bibinfo {year} {2010})},\ \Eprint {http://arxiv.org/abs/0909.2703}
  {arXiv:0909.2703 [cond-mat]} \BibitemShut {NoStop}%
%%CITATION = ARXIV:0909.2703;%%
\bibitem [{\citenamefont {Cortijo}\ and\ \citenamefont
  {Vozmediano}(2007)}]{Cortijo:2006xs}%
  \BibitemOpen
  \bibfield  {author} {\bibinfo {author} {\bibfnamefont {A.}~\bibnamefont
  {Cortijo}}\ and\ \bibinfo {author} {\bibfnamefont {M.~A.~H.}\ \bibnamefont
  {Vozmediano}},\ }\href {\doibase 10.1016/j.nuclphysb.2008.09.006,
  10.1016/j.nuclphysb.2006.10.031} {\bibfield  {journal} {\bibinfo  {journal}
  {Nucl. Phys.}\ }\textbf {\bibinfo {volume} {B763}},\ \bibinfo {pages} {293}
  (\bibinfo {year} {2007})},\ \bibinfo {note} {[Nucl. Phys.B807,659(2009)]},\
  \Eprint {http://arxiv.org/abs/cond-mat/0612374} {arXiv:cond-mat/0612374
  [cond-mat]} \BibitemShut {NoStop}%
%%CITATION = COND-MAT/0612374;%%
\bibitem [{\citenamefont {de~Juan}\ \emph {et~al.}(2010)\citenamefont
  {de~Juan}, \citenamefont {Cortijo},\ and\ \citenamefont
  {Vozmediano}}]{deJuan2010}%
  \BibitemOpen
  \bibfield  {author} {\bibinfo {author} {\bibfnamefont {F.}~\bibnamefont
  {de~Juan}}, \bibinfo {author} {\bibfnamefont {A.}~\bibnamefont {Cortijo}}, \
  and\ \bibinfo {author} {\bibfnamefont {M.~A.~H.}\ \bibnamefont
  {Vozmediano}},\ }\href {\doibase 10.1016/j.nuclphysb.2009.11.012} {\bibfield
  {journal} {\bibinfo  {journal} {Nucl. Phys. B}\ }\textbf {\bibinfo {volume}
  {828}},\ \bibinfo {pages} {625} (\bibinfo {year} {2010})},\ \Eprint
  {http://arxiv.org/abs/0909.4068} {arXiv:0909.4068 [cond-mat]} \BibitemShut
  {NoStop}%
%%CITATION = ARXIV:0909.4068;%%
\bibitem [{\citenamefont {Vozmediano}\ \emph {et~al.}(2010)\citenamefont
  {Vozmediano}, \citenamefont {Katsnelson},\ and\ \citenamefont
  {Guinea}}]{Vozmediano:2010zz}%
  \BibitemOpen
  \bibfield  {author} {\bibinfo {author} {\bibfnamefont {M.~A.~H.}\
  \bibnamefont {Vozmediano}}, \bibinfo {author} {\bibfnamefont {M.~I.}\
  \bibnamefont {Katsnelson}}, \ and\ \bibinfo {author} {\bibfnamefont
  {F.}~\bibnamefont {Guinea}},\ }\href {\doibase 10.1016/j.physrep.2010.07.003}
  {\bibfield  {journal} {\bibinfo  {journal} {Phys. Rept.}\ }\textbf {\bibinfo
  {volume} {496}},\ \bibinfo {pages} {109} (\bibinfo {year}
  {2010})}\BibitemShut {NoStop}%
%%CITATION = PRPLC,496,109;%%
\bibitem [{\citenamefont {Amorim}\ \emph {et~al.}(2016)\citenamefont {Amorim}
  \emph {et~al.}}]{Amorim:2015bga}%
  \BibitemOpen
  \bibfield  {author} {\bibinfo {author} {\bibfnamefont {B.}~\bibnamefont
  {Amorim}} \emph {et~al.},\ }\href {\doibase 10.1016/j.physrep.2015.12.006}
  {\bibfield  {journal} {\bibinfo  {journal} {Phys. Rept.}\ }\textbf {\bibinfo
  {volume} {617}},\ \bibinfo {pages} {1} (\bibinfo {year} {2016})},\ \Eprint
  {http://arxiv.org/abs/1503.00747} {arXiv:1503.00747 [cond-mat]} \BibitemShut
  {NoStop}%
%%CITATION = ARXIV:1503.00747;%%
\bibitem [{\citenamefont {Kibble}(1961)}]{KibbleTorsion1961}%
  \BibitemOpen
  \bibfield  {author} {\bibinfo {author} {\bibfnamefont {T.~W.~B.}\
  \bibnamefont {Kibble}},\ }\href@noop {} {\bibfield  {journal} {\bibinfo
  {journal} {Journal of Mathematical Physics}\ }\textbf {\bibinfo {volume}
  {2}},\ \bibinfo {pages} {212} (\bibinfo {year} {1961})}\BibitemShut {NoStop}%
\bibitem [{\citenamefont {Hehl}\ \emph {et~al.}(1976)\citenamefont {Hehl},
  \citenamefont {Von Der~Heyde}, \citenamefont {Kerlick},\ and\ \citenamefont
  {Nester}}]{Hehl:1976kj}%
  \BibitemOpen
  \bibfield  {author} {\bibinfo {author} {\bibfnamefont {F.~W.}\ \bibnamefont
  {Hehl}}, \bibinfo {author} {\bibfnamefont {P.}~\bibnamefont {Von Der~Heyde}},
  \bibinfo {author} {\bibfnamefont {G.~D.}\ \bibnamefont {Kerlick}}, \ and\
  \bibinfo {author} {\bibfnamefont {J.~M.}\ \bibnamefont {Nester}},\ }\href
  {\doibase 10.1103/RevModPhys.48.393} {\bibfield  {journal} {\bibinfo
  {journal} {Rev. Mod. Phys.}\ }\textbf {\bibinfo {volume} {48}},\ \bibinfo
  {pages} {393} (\bibinfo {year} {1976})}\BibitemShut {NoStop}%
%%CITATION = RMPHA,48,393;%%
\bibitem [{\citenamefont {Hehl}\ and\ \citenamefont
  {Obukhov}(2007)}]{CartanTheoryHehl}%
  \BibitemOpen
  \bibfield  {author} {\bibinfo {author} {\bibfnamefont {F.~W.}\ \bibnamefont
  {Hehl}}\ and\ \bibinfo {author} {\bibfnamefont {Y.~N.}\ \bibnamefont
  {Obukhov}},\ }\href@noop {} {\bibfield  {journal} {\bibinfo  {journal}
  {Annales Fond. Broglie}\ }\textbf {\bibinfo {volume} {32}},\ \bibinfo {pages}
  {157} (\bibinfo {year} {2007})},\ \Eprint {http://arxiv.org/abs/0711.1535}
  {arXiv:0711.1535 [gr-qc]} \BibitemShut {NoStop}%
%%CITATION = ARXIV:0711.1535;%%
\bibitem [{\citenamefont {Pagani}\ and\ \citenamefont
  {Percacci}(2015)}]{PercacciTorsion}%
  \BibitemOpen
  \bibfield  {author} {\bibinfo {author} {\bibfnamefont {C.}~\bibnamefont
  {Pagani}}\ and\ \bibinfo {author} {\bibfnamefont {R.}~\bibnamefont
  {Percacci}},\ }\href {http://stacks.iop.org/0264-9381/32/i=19/a=195019}
  {\bibfield  {journal} {\bibinfo  {journal} {Classical and Quantum Gravity}\
  }\textbf {\bibinfo {volume} {32}},\ \bibinfo {pages} {195019} (\bibinfo
  {year} {2015})}\BibitemShut {NoStop}%
\bibitem [{\citenamefont {Wess}\ and\ \citenamefont
  {Bagger}(1992)}]{WessBagger}%
  \BibitemOpen
  \bibfield  {author} {\bibinfo {author} {\bibfnamefont {J.}~\bibnamefont
  {Wess}}\ and\ \bibinfo {author} {\bibfnamefont {J.}~\bibnamefont {Bagger}},\
  }\bibfield  {booktitle} {\emph {\bibinfo {booktitle} {Supersymmetry and
  Supergravity by Julius Wess and Jonathan Bagger. Princeton University Press,
  2001. ISBN: 978-0-691-02530-8}},\ }\href@noop {} {\ \textbf {\bibinfo
  {volume} {294}} (\bibinfo {year} {1992})}\BibitemShut {NoStop}%
\bibitem [{\citenamefont {Lazar}(2003)}]{Lazar2003}%
  \BibitemOpen
  \bibfield  {author} {\bibinfo {author} {\bibfnamefont {M.}~\bibnamefont
  {Lazar}},\ }\href {\doibase 10.1088/0305-4470/36/5/316} {\bibfield  {journal}
  {\bibinfo  {journal} {J. Phys.}\ }\textbf {\bibinfo {volume} {A36}},\
  \bibinfo {pages} {1415} (\bibinfo {year} {2003})},\ \Eprint
  {http://arxiv.org/abs/cond-mat/0208360} {arXiv:cond-mat/0208360 [cond-mat]}
  \BibitemShut {NoStop}%
%%CITATION = COND-MAT/0208360;%%
\end{thebibliography}%

\end{document}